\documentclass[fleqn,usenatbib]{mnras}

\usepackage{mathptmx}

\usepackage[T1]{fontenc}

\DeclareRobustCommand{\VAN}[3]{#2}
\let\VANthebibliography\thebibliography
\def\thebibliography{\DeclareRobustCommand{\VAN}[3]{##3}\VANthebibliography}

\usepackage{graphicx}	
\usepackage{amsmath}	
\usepackage{amssymb}	
\usepackage{enumitem}
\setlist{itemsep=2mm}

\newcommand{\MYhref}[3][blue]{\href{#2}{\color{#1}{#3}}}%

\title[A precise photometric ratio I]{A precise photometric ratio via laser excitation of the sodium layer -- I. \\
                                      {\fontsize{14}{16.8}\selectfont One-photon excitation using 342.78~nm light}}

\author[J. E. Albert et al.]{Justin E. Albert \href{https://orcid.org/0000-0003-0253-2505}{\includegraphics[scale=0.55]{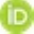}},$^{1}$\thanks{E-mail: \href{mailto:jalbert@uvic.ca}{jalbert@uvic.ca} (JEA)}
Dmitry Budker \href{https://orcid.org/0000-0002-7356-4814}{\includegraphics[scale=0.55]{orcid}},$^{2,4,5}$
Kelly Chance \href{https://orcid.org/0000-0002-7339-7577}{\includegraphics[scale=0.55]{orcid}},$^{3}$
Iouli E. Gordon \href{https://orcid.org/0000-0003-4763-2841}{\includegraphics[scale=0.55]{orcid}},$^{3}$\newauthor
Felipe Pedreros Bustos \href{https://orcid.org/0000-0001-7664-1590}{\includegraphics[scale=0.55]{orcid}},$^{2,6}$
Maxim Pospelov,$^{1,7,\ddagger}$
Simon M. Rochester \href{https://orcid.org/0000-0001-5202-5718}{\includegraphics[scale=0.55]{orcid}}\hspace*{0.35mm}$^{4}$
and H. R. Sadeghpour \href{https://orcid.org/0000-0001-5707-8675}{\includegraphics[scale=0.55]{orcid}}\hspace*{0.35mm}$^{3}$
\\
$^{1}$Department of Physics and Astronomy, University of Victoria, Victoria, British Columbia V8W 3P6, Canada\\
$^{2}$Helmholtz Institute, Johannes Gutenberg-Universit\"at Mainz, 55099 Mainz, Germany\\
$^{3}$ITAMP, Center for Astrophysics $\vert$ Harvard \& Smithsonian, Cambridge, Massachusetts 02138, USA\\
$^{4}$Rochester Scientific LLC, El Cerrito, California 94530, USA\\
$^{5}$Department of Physics, University of California, Berkeley, California 94720-7300, USA\\
$^{6}$Laboratoire d'Astrophysique de Marseille (LAM), Universit\'e d'Aix-Marseille \& CNRS, F-13388 Marseille, France\\
$^{7}$Perimeter Institute of Theoretical Physics, Waterloo, Ontario N2L 2Y5, Canada\\
$^{\ddagger}$Now at School of Physics and Astronomy, University of Minnesota, Minneapolis, Minnesota 55455, USA
\vspace*{-3mm}
}

\date{Accepted 2021 June 3. Received 2021 June 3; in original form 2021 January 18}

\pubyear{2021}

\begin{document}
\jname{\MYhref[magenta]{https://doi.org/10.1093/mnras/stab1621}{\textbf{MNRAS}}}
\volume{508}
\label{firstpage}
\pagerange{\MYhref[blue]{https://doi.org/10.1093/mnras/stab1621}{4399--4411}}
\maketitle

\begin{abstract}
\hyphenpenalty 10000
\exhyphenpenalty 10000
The largest uncertainty on measurements of dark energy using type~Ia supernovae is presently due to
systematics from photometry; specifically to the relative uncertainty on photometry as a
function of wavelength in the optical spectrum.
We show that a precise constraint on relative photometry between the visible and near-infrared can be
achieved at upcoming survey telescopes, such as
at the Vera C.~Rubin Observatory,
via a 
laser source tuned to
the 342.78~nm vacuum excitation wavelength of
neutral sodium atoms.
Using a high-power 
laser,
this excitation will produce an artificial star,
which we term a ``laser photometric ratio star'' (LPRS) of de-excitation light in the mesosphere 
at wavelengths in vacuum of
589.16~nm, 589.76~nm, 818.55~nm, and 819.70~nm, with the sum of the numbers of 589.16~nm and 589.76~nm
photons produced by this process
equal to the sum of the numbers of 818.55~nm and 819.70~nm photons,
establishing a precise calibration ratio between, for
example, the $r$ and $z$ filters of the LSST camera at the Rubin Observatory.
This technique can thus provide a novel mechanism for establishing a
spectrophotometric calibration ratio of unprecedented precision
for upcoming telescopic observations across astronomy and atmospheric physics;
thus greatly improving the performance of upcoming measurements of dark energy parameters using type~Ia supernovae.
The second paper of this pair describes an alternative
technique to achieve a similar, but brighter, LPRS than the technique described in this paper, by using two 
lasers
near
resonances at 
589.16~nm and 819.71~nm,
rather than the single 342.78~nm on-resonance laser technique described in this paper.
\end{abstract}




\section{Motivations}

Over two-thirds of the total mass-energy of the Universe is
dark energy, a mysterious
characteristic of space
with measured properties that are consistent, at present, with being those of the cosmological constant from the field
equations of general relativity~\citep{Planck2020, PDGRevCosmParams}.  The value of the cosmological constant, and its
possible relation to the zero-point energies of quantum fields is, however, a notoriously longstanding problem at
the intersection of quantum mechanics and general relativity that predates the discovery, two decades ago, of
non-zero dark energy~\citep{HighZ, SCP} by several additional decades~\citep{Weinberg89}.
The large uncertainties at present
on multiple parameters of dark energy, especially on any changes that dark energy may have undergone over cosmic
history, ensure that the improvement in measurement of dark energy properties remains at the forefront of observational
cosmology and astrophysics.

\begin{figure*}
\begin{center}
\hypertarget{fig1}{
\includegraphics[scale=.7]{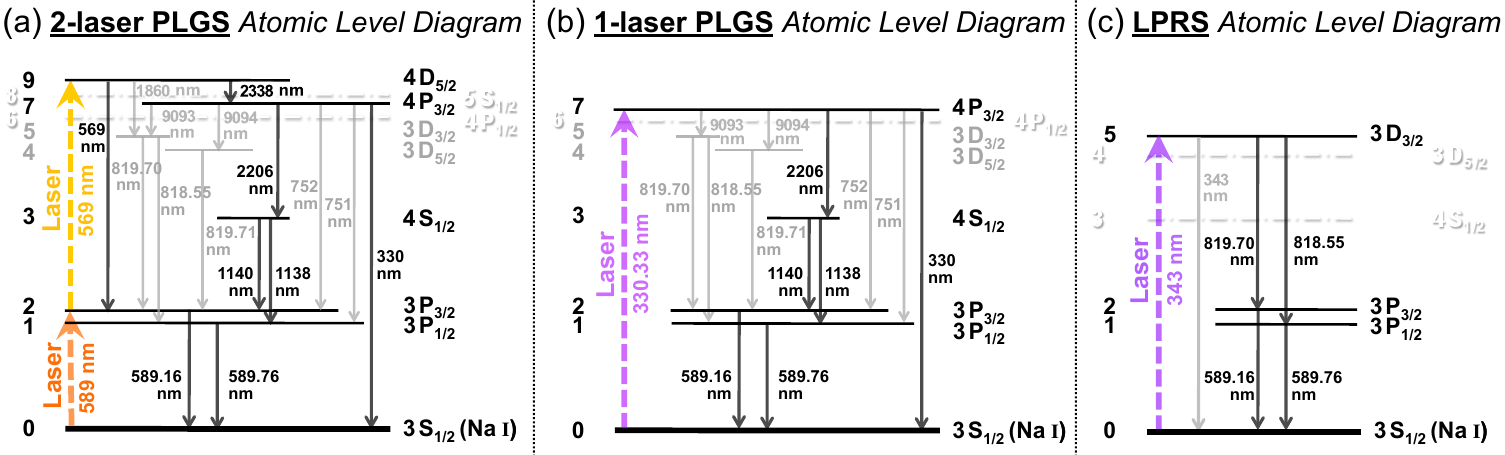}
}
\end{center}
\caption{Atomic level diagrams for neutral sodium atoms (not to scale): (a)~for two-laser polychromatic laser guide stars (PLGS),
(b)~for one-laser PLGS, and (c)~for a laser photometric ratio star (LPRS).  ``Forbidden'' magnetic dipole and electric
quadrupole transitions, and also the levels (and decays from the levels) that are only accessible by such transitions,
are in solid gray, whereas
allowed and laser-excited transitions and levels are in solid black.
``Ghost'' levels, that are entirely inaccessible from states in the diagrams that are excited by the lasers, are shown in
shadowed gray text and dash-dotted lines.  As can be seen, the relatively simple LPRS diagram results in ``fully-mandated cascades'' from
the 819/820~nm de-excitations to the 589/590~nm de-excitations, resulting in a mandated 1:1 ratio between those
produced photons, whereas the cascades in the relatively more complex PLGS diagrams do not have such
absolutely mandated ratios between their produced photons of different wavelengths.
(The simplicity of the LPRS diagram is counterbalanced by the weakness of the 343~nm
transition when compared with the transitions excited by PLGS lasers, which results in LPRS being much dimmer than
PLGS.)}
\label{fig:PLGSLPRSEnergyDiagrams}
\end{figure*}

Measurements of dark energy utilizing the method of its discovery, i.e.~the construction of a Hubble curve with
type~Ia supernovae (SNeIa) have, for the past ten years or so, been limited primarily
by systematic uncertainty on the measurement of astronomical magnitude as a function of colour
within the optical spectrum~\citep{Betoule14, WoodVasey07}.  The statistics of observed SNeIa have continued to
improve~\citep{Jones18}, and will improve dramatically with the beginning of the Vera C.~Rubin Observatory
sky survey (known as the ``Legacy Survey of Space and Time,'' or LSST) in the early 2020s~\citep{LSSTOverview}.
The systematic limitation from spectrophotometry will continue, however, to be a fundamental barrier to major improvement
on measurements in this area~\citep{StuTon06}, in the absence of novel technology for spectrophotometric calibration with 
a precision corresponding to the continuing dramatic improvement in observed SNeIa statistics.

Beyond dark energy measurements with SNeIa, the precision of relative spectrophotometry also is presently a limiting
factor in measurements of stellar populations in galaxy clusters~\citep{Connor17} and in upcoming photometric redshift
surveys measuring growth of structure~\citep{Kirk15}.

\section{Existing Related Infrastructure and the Atmospheric Sodium Layer}
\label{sec:ExtInfAndNaLayer}

Within the separate astronomical domain of point-spread function (PSF) minimization and calibration, laser guide stars (LGS) have
dramatically improved the angular resolution of ground-based telescopic imaging and spectroscopy over the past decades,
within the fields of view of major deep-field observatories employing LGS and adaptive optics (AO)~\citep{Wiz06,Bon02,OliMax94}.
LGS produce resonant scattering of light from the layer of atomic sodium in the Earth's mesosphere
by utilizing lasers located at observatory sites and typically tuned to the sodium D$_2$-line resonance at 589~nm.

The Earth's atmospheric layer of neutral atomic sodium (\ion{Na}{i})~\citep{Slipher29} exists between approximately 80 and 105~km above 
sea level.  It originates primarily from the ablation of meteors in the ionosphere~\citep{Chapman39,Pla15}; with a potentially significant 
fraction of upper-atmospheric \ion{Na}{i} atoms having instead been (somewhat surprisingly) previously located on the regolith of the Moon,
ejected from the Moon's surface by meteoritic impacts, swept into a long lunar \ion{Na}{i} tail that is pointed away from the Sun, 
and then intercepted by the Earth during new moon periods~\citep{PotMor88b,PotMor88a,Bau21}.
The Earth's total atmospheric column density of sodium
varies with time and location
between about $2 \times 10^{9}$~atoms/cm$^2$ and $8 \times
10^{9}$~atoms/cm$^2$~\citep{Megie78,Mou10}.
Current LGS typically use a solid-state laser or a fiber laser,\footnote{Historically, most LGS lasers were dye-based.} with typical optical output power around 
10~--~20~W,
directed into the mesosphere to produce an artificial star at about $9^{th}$~magnitude,\footnote{Typical LGS return flux is in the range
(5~--~25)$\times 10^6$~photons/s/m$^{2}$.} which is sufficient for real-time deformable mirror AO image
correction at 4~m-class and larger observatories.

Sodium is not especially abundant in comparison to other atomic and molecular species in the upper atmosphere, but the product of its density and
its optical cross-section [with Einstein $A$ coefficients of $6.16 \times 10^7$ s$^{-1}$
and $6.14 \times 10^7$ s$^{-1}$~\citep{Jun81,NISTtables} for the Na 589.16 and 589.76 nm resonances\footnote{In this paper, wavelengths are given
in vacuum, typically to either the nearest nanometre, or nearest hundredth of a nanometre.  Wavelengths for sodium are as provided by~\citet{Kel08}.}
respectively]
makes sodium the most favorable element for optical excitation.\footnote{Other elements with weakly-bound outer-shell electrons,
such as potassium (which also has strong D-line resonances, at
767~nm and 770~nm), and calcium (with a strong resonance at 423~nm), have optical transitions of similar strength to the Na D lines, but
the paucity of those elements in the Earth's upper atmosphere, relative to Na, reduces their potential for LGS utilization.}
Due to the poorly-predictable variability of the Na column density (as well as other trace elemental column densities),
LGS do not have major utility for precise calibration of photometry,
but rather find their utility, as mentioned above, in minimization and calibration of PSF, when combined with AO.

\begin{figure*}
\begin{center}
\hypertarget{fig2}{
\includegraphics[scale=.59]{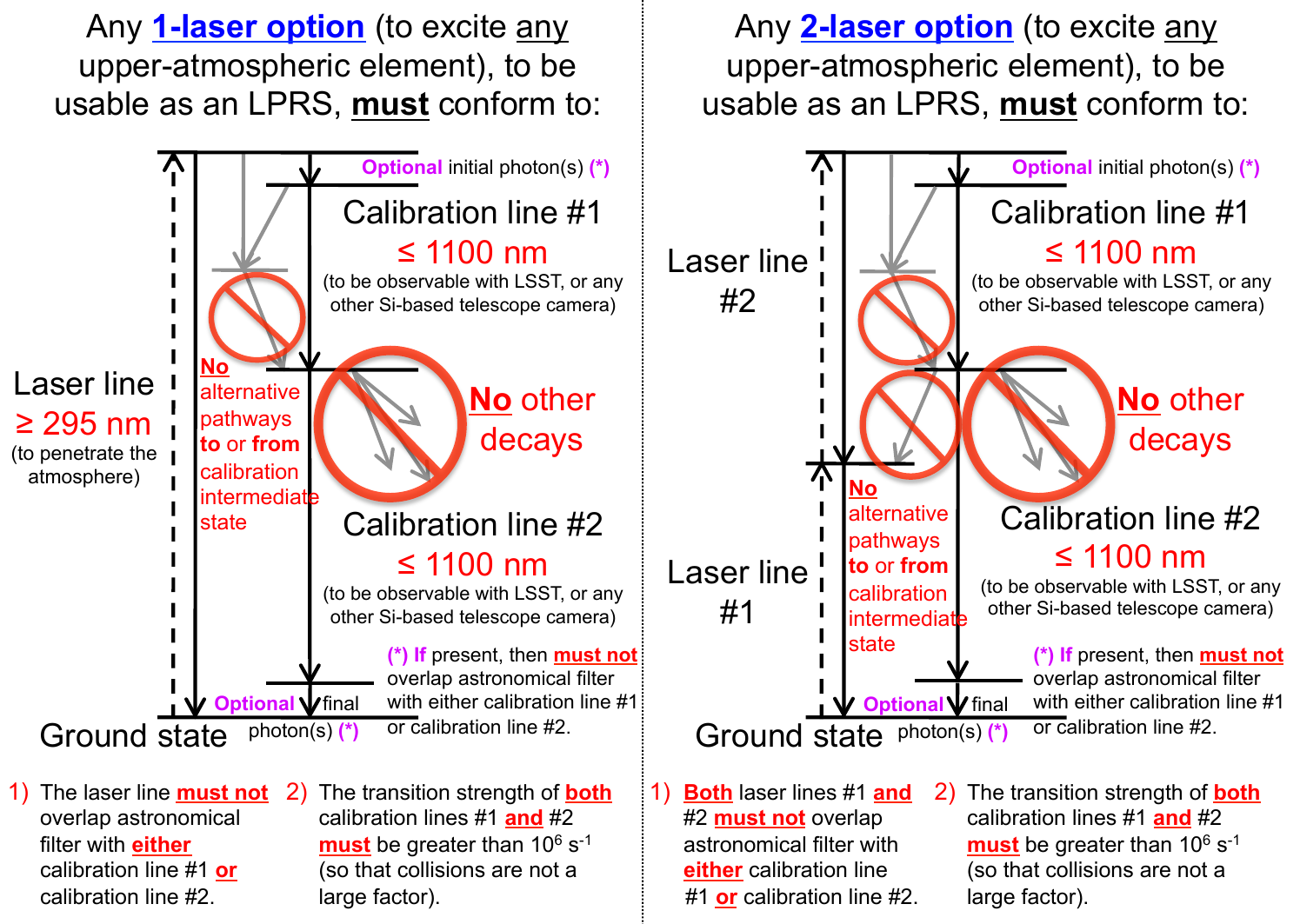}
}
\end{center}
\caption{Constraints on atomic  
level diagrams for a laser photometric ratio star (LPRS) in the case of one-laser on-resonance excitation (left), and
two-laser on-resonance excitation (right).}
\label{fig:LPRSConstraints}
\end{figure*}

Polychromatic LGS (PLGS), which produce artificial stars with one or more optical wavelengths in addition to either
one or both of the sodium D lines, were first conceived by~\citet{Foy95} and have since been tested on the
sky~\citep{Foy00}, but not yet utilised within a full closed AO loop.
PLGS have a significant advantage over monochromatic LGS (MLGS) in that PLGS
can, in principle, provide the information necessary to compensate first-order atmospheric aberrations
(``tip-tilt'') based on differential measurements of the tip-tilt at two separated wavelengths~\citep{Tys15,FoyPiq},
whereas MLGS lack this capability (necessitating the simultaneous use of a sufficiently-bright natural guide star, which has a low probability
of existing in close vicinity to the object of interest).  The PLGS
use two lasers, at 589 nm and 569 nm, producing an artificial star with wavelengths as shown in Fig.~\ref{fig:PLGSLPRSEnergyDiagrams}(a).
An alternative conceptual PLGS utilizing a single laser tuned to the $4\,$P$_{3/2}$ Na excitation
at 330 nm has also been proposed~\citep{Piq06} and tested in the laboratory~\citep{Mol07,Che16}, to produce an artificial star with
wavelengths as shown in Fig.~\ref{fig:PLGSLPRSEnergyDiagrams}(b).  As shown in Fig.~\ref{fig:PLGSLPRSEnergyDiagrams}, however, the
production ratios between photons of different wavelengths from either 2-laser or 1-laser PLGS are not absolutely mandated by
a single direct cascade
but, rather, depend on the ratios of different transition strengths and, thus, on uncertainties in the precise knowledge of those
transition strengths; whereas the laser photometric ratio star (LPRS) described in the next section
and in Fig.~\ref{fig:PLGSLPRSEnergyDiagrams}(c) does not have that limitation.  Additionally, an LPRS will produce all of its calibration photons
at wavelengths less than the
$\sim$1100~nm maximum wavelength that would be observable with the LSST camera at the Rubin Observatory, and other 
silicon-based optical cameras; whereas all 1138~nm and above 
photons produced by PLGS are above the maximum Si-observable wavelength.\footnote{Nevertheless (even if neither 1:1, nor presently known with
high precision), the {\it temporal stability} of the production ratio of 330~nm photons to 589/590~nm photons from a 1-laser PLGS might have
some utility for relative photometric calibration between 330~nm and 589/590~nm at observatories; although the time-dependence of
near-180\degr~Rayleigh scattering in the lower atmosphere of 330~nm photons from the laser could confound such a photometric-ratio usage of
a 1-laser PLGS.}

\section{Concept: Principle of Laser Photometric Ratio Stars (LPRS)}

As shown in Fig.~\ref{fig:PLGSLPRSEnergyDiagrams}(c), the $3\,$D$_{3/2}$ state of neutral sodium atoms can be photoexcited 
by a 342.78~nm laser, resulting in a ``fully-mandated cascade'' of two photons: a 819/820~nm photon followed by
a 589/590~nm photon.\footnote{The only other de-excitation option from the $3\,$D$_{3/2}$ state is the
emission of a single photon of the same 342.78~nm wavelength that initially excited the atom.
That is an electric quadrupole transition that is suppressed compared to the 819/820~nm photon + 589/590~nm photon
channel, and also does not affect the 1:1 ratio between 819/820~nm photon and 589/590~nm photon production.}
This mandates a 1:1 ratio between production of 819/820~nm photons vs.~589/590~nm photons.  We importantly 
note that the 342.78~nm transition is a ``forbidden'' electric quadrupole transition with Einstein $A$ coefficient
within\footnote{\label{foot:EinA}The precise value of this Einstein $A$ coefficient has been subject to some controversy:
\citet{Her69} experimentally measured the value $(6.9 \pm 0.7) \times 10^2$~s$^{-1}$.  A theoretical calculation
was then performed by~\citet{Ali71} which agreed with that experimental result.  However, a ``substantial error''
in the~\citet{Ali71} calculation was noticed by~\citet{McE73}, and the revised theoretical estimate of $2.1 \times 10^2$~s$^{-1}$
resulted, which is just over 3$\times$ smaller than the~\citet{Her69} experimental measurement.  That discrepancy has not yet been resolved with
additional measurements or calculations.  We conservatively use the lower
$2.1 \times 10^2$~s$^{-1}$~\citet{McE73} theoretical estimate throughout this paper.}
the range (2 -- 7)$\times 10^2$~s$^{-1}$, approximately five orders of magnitude smaller than the (precisely
measured) $6.2 \times 10^7$~s$^{-1}$, $1.2 \times 10^7$~s$^{-1}$, and $2.7 \times 10^6$~s$^{-1}$ Einstein $A$ coefficients
for the allowed 589~nm, 569~nm, and 330~nm transitions respectively~\citep{Jun81,Mei37,NISTtables}.
Although the 342.78~nm transition is, thus, much weaker than the 589~nm, 569~nm, or 330~nm transitions that are
excited by PLGS lasers, with the use of a powerful ($\ge 500$~W average power) laser,
342.78~nm is still definitely a strong enough transition to produce an LPRS
for use at the Rubin Observatory and other 8~m-class (or larger) telescopes,
as calculated below in Sections~\ref{sec:EstFlux}~--~\ref{sec:EstImp}.

\section{Other Upper-Atmospheric Excitations Considered}

To search for alternative
artificial sources in the upper atmosphere with precise photometric ratios in the optical
domain, that would also be useful for performing relative
photometric calibration of the combined throughput of atmosphere, telescope, and camera between
wavelengths in different optical filters, we considered
various different possible sets of atmospheric atomic and molecular
optical excitations.  Figure~\ref{fig:LPRSConstraints} shows constraints on properties of atomic systems
that could be useful for such calibrations, in both one-laser and two-laser on-resonance excitation schemes.

\defcitealias{Alb21a}{Paper~II}

We developed a code,
{\tt LPRSAtomicCascadeFinder},\footnote{Available from the authors upon request.}
to search the~\citet{NISTtables}
database for sets of atomic transitions that would obey the constraints shown in the Fig.~\ref{fig:LPRSConstraints}
diagrams.  In addition to neutral sodium (\ion{Na}{i}), we ran {\tt LPRSAtomicCascadeFinder} on the \ion{Al}{i}, 
\ion{C}{i}, \ion{Ca}{i}, \ion{Fe}{i}, \ion{H}{i}, \ion{He}{i}, \ion{K}{i}, \ion{N}{i}, \ion{Ne}{i}, \ion{O}{i}, \ion{Al}{ii},
\ion{C}{ii},  \ion{Ca}{ii}, \ion{Fe}{ii}, \ion{H}{ii}, \ion{He}{ii}, \ion{K}{ii}, \ion{N}{ii}, \ion{Na}{ii}, \ion{Ne}{ii}, and \ion{O}{ii} tables
from~\citet{NISTtables}.  The only set of atomic transitions
found is the 342.78~nm excitation of
\ion{Na}{i} [shown in Fig.~\ref{fig:PLGSLPRSEnergyDiagrams}(c) and described in the previous section.
However, please see~\citet{Alb21a} (hereafter referred to as~\citetalias{Alb21a}) for two-laser \textit{off-resonance}
atomic excitation options for LPRS generation.]

\begin{figure}
\begin{center}
\includegraphics[scale=.38]{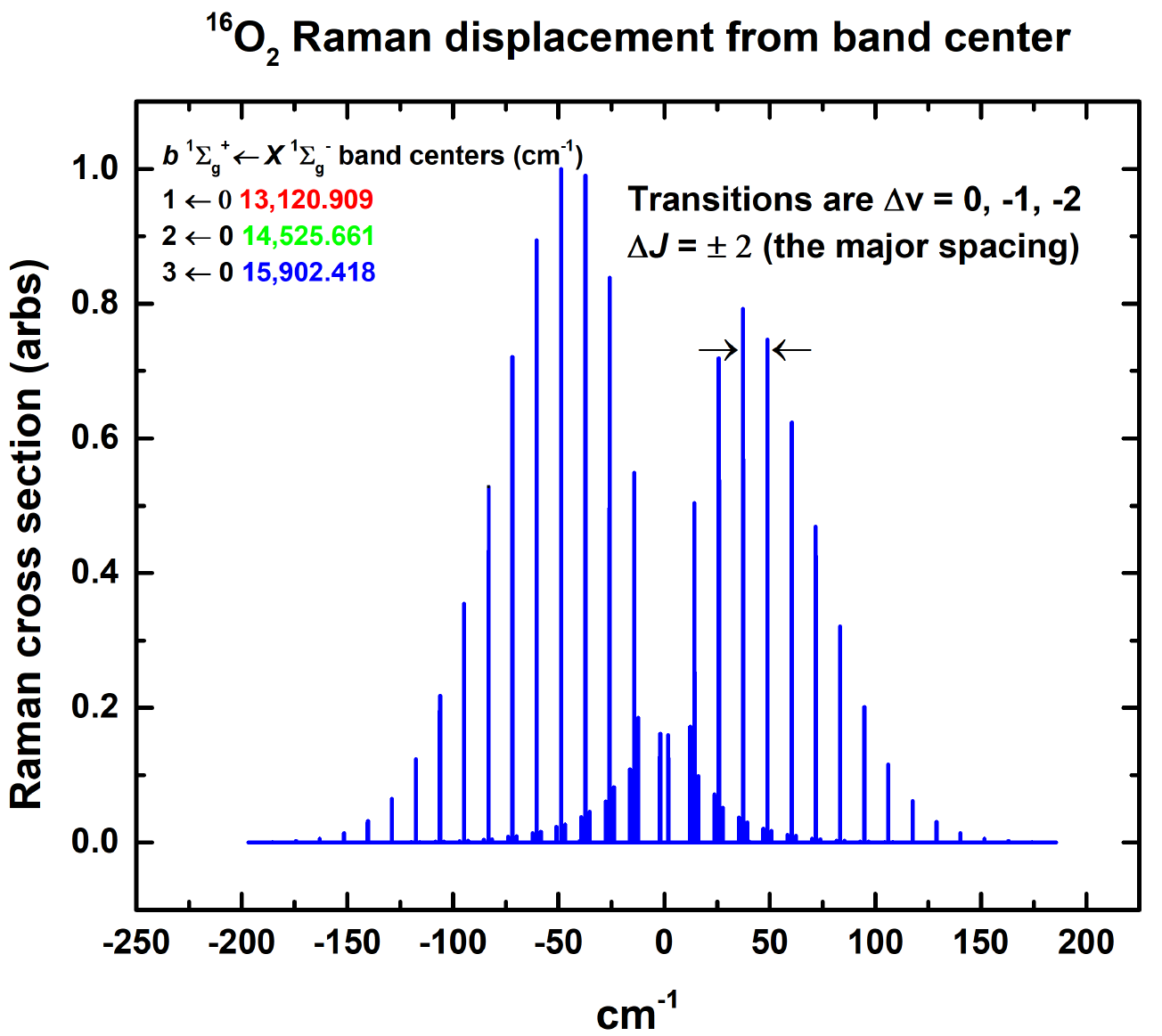} 
\end{center}  
\vspace*{-0.3cm}
\caption{Spectrum of resonant Raman scattering in $^{16}$O$_2$ in the Schuman-Runge electronic bands [data from~\citet{HITRAN16}].}
\label{fig:ResRamanEdited}
\end{figure}

In regard to
molecular excitations, there are multiple compounds in both the lower and upper atmosphere (for example $^{16}$O$_2$, as shown in
Fig.~\ref{fig:ResRamanEdited}) with spectra with relative wavelengths that are known to extraordinarily high precision
(e.g.~to a part in $10^7$, or even better in some cases).
Recently, by using Raman spectroscopy of multiple atmospheric components with scattered
light returned from the laser guide star system of the Very Large Telescope (VLT) at Paranal in Chile,
precise calibration of the VLT ESPRESSO spectrograph was achieved~\citep{Vog19}.
However, the relative cross-sections corresponding to the different
spectral lines --- which are the quantities that are important for relative \textit{photometric}, rather than \textit{spectroscopic}, calibration ---
are known to far lower precisions: typically of order (5 -- 20)\%~\citep{Sim97,Bul96}, and thus unfortunately would
not approach the precision of relative photometric calibration using the far more precisely-predicted 1:1 photometric ratio of 819/820~nm
photons to 589/590~nm photons from 342.78~nm LPRS excitation of the sodium layer.

\section{Laser and Launch Telescope}
\label{sec:Laser}

To drive the weak 342.78~nm dipole parity-forbidden transition in neutral sodium, as shown below in Sections~\ref{sec:EstFlux}~--~\ref{sec:EstImp},
a powerful ($\ge$500~W average power) laser is required to generate an observable and usable
photometric ratio star, even for large telescopes with deep limiting magnitudes, such as at the
Rubin Observatory.\footnote{Per~\citet{LSSTOverview}, the minimum 
single-visit limiting
magnitudes for astronomical point sources using the LSST $ugrizy$ filters 
will be 23.4, 24.6, 24.3, 23.6, 22.9, 21.7 respectively, with exposure times per 
visit of 30~s.}
A 500~W average power 342.78~nm laser with launch telescope is, nevertheless,
clearly technologically achievable.  In this Section, we provide an example set of specifications and a design outline
(consisting of two laser design options)
for a laser and launch telescope that would meet the
requirements for an LPRS for precision photometric calibration for the case of LSST at the Rubin Observatory.

Since the natural linewidth of the 342.78~nm \ion{Na}{i} transition is narrow in comparison with the line broadening induced by the thermal
motion of the \ion{Na}{i} atoms, the maximum optimal laser linewidth will be governed by this Doppler broadening of the transition in the mesosphere:
\begin{equation}
\qquad \qquad \sigma_{\nu} \; \le \; \frac{\nu_0}{c}\sqrt{\frac{k_B T_{\rm 100\:km}}{M_{\rm Na}}} \quad \; {\rm (Doppler \; broadening)},
\end{equation}
where $T_{\rm 100\:km} \approx 200$~K, and will thus be
$\sigma_{\nu} \lesssim 1.5$~GHz (i.e. $\sigma_{\nu} \lesssim 0.05$~cm$^{-1}$, $\sigma_{\lambda} \lesssim 0.0006$~nm).

Solid-state laser systems generally are preferable to dye-based systems due to their higher wall-plug efficiencies, lower maintenance requirements, 
and reduced sensitivity to vibrations and ground movement; however solid-state systems are typically less wavelength-flexible than dye systems, and 
often require longer design times and can have higher design and construction costs.  Advances in optical parametric oscillator (OPO) and in 
closely-related nonlinear optical crystal technology for wavelength flexibility for solid-state lasers, as well as in the output power and linewidth 
of solid-state near-infrared (NIR) pump lasers (including Yb-doped optical fiber- and disk-based pump laser amplification systems), have largely 
amelioriated such disadvantages of solid-state systems.  Here, we provide two options for solid-state lasers that could meet the challenging average 
power, linewidth, and wavelength requirements for creating an LPRS for the Rubin Observatory.

The first option uses a 532~nm single-mode quasi-continuous-wave ($\sim$250~MHz repetition rate) frequency-doubled Nd:YAG fiber laser, with an 
average output power of 1~kW and linewidth of $<\!1$~GHz, from~\citet{IPG} (product VLR-532-1000), as input to a beamsplitter, with one half of the 
resulting light entering a lithium triborate (LBO) OPO crystal for the production of 963.1~nm light (as well as unused 1188.5~nm light); and this 
resulting 963.1~nm light, as well as the other half of the 532~nm light, entering a second LBO crystal that acts as a sum-frequency generator (SFG) 
that combines the 963.1~nm and 532~nm input photons into 342.7~nm output light.  The wavelength of the output light would then be variable between 
342.6~nm and 342.8~nm by a small adjustment of the angle of the OPO crystal.  The conversion efficiencies of the OPO and SFG processes can each 
exceed 50\%.  Although an optical output power of 500~W at 342.7~nm when using an input power of 1~kW at 532 nm would be difficult to achieve, it 
would be both possible and made easier by the likely future increases in available input power from single-mode 532~nm pump lasers (from IPG and 
other suppliers).  The cooling of the OPO and SFG crystals would also be an engineering challenge with such high average optical input powers, but 
would not be insurmountable with a carefully-designed water cooling system for those optics.

The second option would also use two LBO crystals, also for OPO and SFG processes respectively, for the production of 342.7~nm output light, but would 
use as input a 515~nm single-mode frequency-doubled 1030~nm Yb-fiber or Yb:YAG disk pump laser (instead of using 532~nm input).  High-power 1030~nm Yb 
pump lasers are available from, for example, \citet{TRUMPF} (TruDisk product line, which could be made into single-mode lasers) and~\citet{IPG} (product 
YLR-1030-1000), which could then be frequency-doubled to 515~nm output with the standard addition of a high-efficiency second-harmonic generation (SHG) 
crystal.  Following the generation of the 515~nm output, the light path would then proceed in a qualitatively similar way to the first option, however 
with an increase in efficiency from the OPO process, since the necessarily-wasted idler beam would be at a longer wavelength (2071~nm, rather than the 
1188.5~nm in the first option), and thus consume less energy.  Variability of the output wavelength between 342.6~nm and 342.8~nm would also proceed in a 
similar way as in the first option.

Additional conceivable possibilities for producing light with our necessary average output power (${\ge\!500}$~W), linewidth (${\le\!1.5}$~GHz), and 
wavelength (variable between 342.6~nm and 342.8~nm) specifications may include: a) The frequency-tripling (using a third-harmonic generation (THG) 
crystal) of high-power single-mode input light that could vary in wavelength between 1027.8~nm and 1028.4~nm (i.e. just below the 1030~nm 
optimal output wavelength of Yb-fiber or Yb:YAG disk lasers, which could possibly be achieved via a small real-time variation of the properties, such as 
tension or temperature, of the fiber Bragg grating or the Yb:YAG disk); or b) The use of a high-output-power frequency-doubled dye laser (rather than the 
solid-state options above).  However, we believe the two options provided in the above paragraphs would be more practical and relatively easier to 
achieve than either of those possibilities.

The output laser light would be directed into the sky via a low-divergence launch telescope, which
expands the beam, and correspondingly lowers its angular divergence,
in order to minimise the resulting beam diameter at 100 km altitude.  The launch telescope would have the same general optical
design as typical launch telescopes for LGS, i.e.~expansion of the beam to approximately 0.5~m diameter with the minimum achievable wavefront 
error, however the optical elements (lens material and mirror coatings) would of course be optimised for 342.8~nm, rather than for 589~nm as in LGS.   
(Specifically, mirror coatings would be UV-enhanced aluminum\footnote{I.e.~similar, for example, to Thorlabs K07 coating~\citep{Thorlabs},  
which is optimised for light near 350~nm and can achieve reflectance of $>$99\% at 0\degr~incidence angle and $>$98\% at 45\degr~incidence
angle, at wavelengths between 342.6~nm and 342.8~nm.}; and
lens materials would be UV-grade fused silica, MgF$_2$, or CaF$_2$, rather than glass.)  As the laser input to the launch telescope can
achieve a beam quality that is within a factor of 2 of diffraction limitation, the resulting
output beam from the launch telescope can achieve an angular divergence that is below 0.2\arcsec~(the pixel scale of the LSST camera at 
the Rubin Observatory).\footnote{We note that the safety and regulatory aspects of a 500~W laser beam that is outside the visible spectrum would be important.  
Although the beam would of course be directed into the sky and away from humans, the power of the beam and its resulting Rayleigh-scattered light 
could potentially cause eye damage or even burned skin if one were physically too close to the beam path, or were to look directly at it from 
nearby, without any intervening absorber of near-UV light.  We estimate that, if near an observatory while such a beam is on, a distance of 50~m
from the beam if not looking directly at its path, or at least 500~m away if looking directly at its path without eye protection, will be
sufficient for human safety.  To prevent humans from entering the open observatory dome when the beam is on, interlocks would clearly need to be
placed on doors to the dome.  A low-power eye-visible laser beam co-axial with the 342.78~nm beam, with an additional audible alarm both prior
to and during beam turn-on, would give proper warning.  Birds and other wildlife that happened to fly within the path of the beam could be seriously
injured or killed, however fortunately very few birds fly over the top of high-altitude observatory sites at night.  Aircraft or satellites that
happened to pass through the path of the laser beam would do so extremely quickly and thus absorb very little energy on any given surface, so we
view that as less dangerous than humans being outdoors in the near vicinity of the observatory when such a beam is operational.}

\setcounter{subsection}{1}

\subsubsection{LPRS size and ellipticity} The resulting diameter of the beam at the 100~km altitude of the sodium layer will
approximately equal the sum in quadrature of the beam diameter
at launch telescope exit, the expansion of the beam in the atmosphere due to its angular divergence at launch telescope exit, and the
expansion of the beam in the atmosphere due to angular divergence caused by atmospheric turbulence.  In clear conditions, total atmospheric
divergence in a vertical path due to typical amounts of turbulence is at the level of approximately
5~$\mu$rad $\approx$ 1\arcsec~\citep{Tatarski}, and thus the beam diameter at 100~km would be, thus, approximately
$\sqrt{(0.5)^2 + (0.1)^2 + (0.5)^2}$~m $\approx 0.7$~m, i.e.~about 1.4\arcsec~on the sky.

A small additional enlargement of the LPRS beam diameter in a radial direction outward from the centre of the telescopic field of view would
occur, for the reason that, as shown in Fig.~\ref{fig:KeckRayleigh}(a), the centre of the laser launch telescope would be slightly offset from the
centre of the aperture of the observing telescope.  That, together with the finite extent of the sodium layer in Earth's atmosphere, would
result in an additional angular diameter of the LPRS spot, corresponding to the angular diameter of the
linear intersection of the laser beam with the vertical extent of the sodium layer.
(This effect would be similar and completely analogous to observed ellipticity of LGS,
for an identical reason.)  The LPRS would thus be approximately elliptical in shape, rather than perfectly circular, on the field of view.
We calculate that the resulting eccentricity of the LPRS ellipse would be approximately 0.75 (i.e. that the major axis diameter of
the LPRS ellipse will be approximately 2.1\arcsec~on the sky, with the minor axis diameter being the previously-calculated 1.4\arcsec),
under simple assumptions that the LPRS beam is launched 4 
metres offset from the centre of the telescope aperture,
and that the number density of \ion{Na}{i} atoms is approximately Gaussian-distributed and centered at 90~km above the elevation of the telescope,
with 102.5~km and 77.5~km above telescope elevation respectively forming the upper and lower 90\% confidence level limits of this Gaussian
distribution (i.e. that the standard deviation of the vertical distribution of \ion{Na}{i} number density is approximately 8~km).

In the absence of atmospheric turbulence, and with an upward-launched laser beam having a perfectly spatially Gaussian-distributed
intensity profile, the LPRS profile would then, in that case --- together with the above-mentioned number density of
\ion{Na}{i} atoms being hypothetically Gaussian-distributed in altitude --- be a perfectly Gaussian ellipse on the sky.  As shown 
in, for example, \citet{Hol10}, real LGS profiles on the sky, and by extension the profile of an LPRS on the sky, will have 
larger tails and be more complex to precisely parametrize than would a simple 2-dimensional Gaussian-distributed spatial profile.
However, for the sake of simplicity, we make the assumption in the analysis in Sections~\ref{sec:EstFlux}~--~\ref{sec:EstImp} of this
paper that the LPRS profile on the sky will be a perfectly Gaussian ellipse.  Although this will likely tend to slightly
overestimate the observed signal flux, and thus slightly underestimate the resulting uncertainty on the observed photometric ratio,
we feel that the resulting corrections from these effects to the analysis performed in this paper are likely to be small.

\subsubsection{Flat-fielding} Uncertainties related to the flat-fielding of photometric calibration data across the full
focal plane can also constitute a significant
component of systematic uncertainty for SNeIa dark energy and other measurements~\citep{Jones18,Betoule14,WoodVasey07}.  With the
laser launch telescope rigidly affixed to the main telescope outer support structure as in Fig.~\ref{fig:KeckRayleigh}(a), the resulting
LPRS would be in a fixed location on the focal plane, and thus the LPRS would not address such uncertainties related to flat-fielding.
However, if the launch telescope were instead mounted to the main telescope structure on a tip-tilt mount, so that the launch telescope
could slightly tip and tilt in altitude and azimuth with respect to the main telescope (up to approximately a degree in the two directions),
the LPRS could then be moved around the focal plane as needed in order to ameliorate flat-fielding uncertainties.

\subsubsection{Intra-filter relative spectrophotometry} The LPRS techniques described in this paper, as well as in~\citetalias{Alb21a},
can provide precise int\textit{er}-filter calibration constants to quantify the
total relative throughput of atmosphere, telescope, camera, and detector at 589/590~nm wavelengths (at which the $r$ filter would be used)
vs.~at 819/820~nm wavelengths (at which the $i$ or $z$ filter would be used).  Int\textit{ra}-filter calibration constants, to quantify the
total throughput ratio between pairs of wavelengths within the bandpass of any single filter (or between any pairs of filters other than
$r$ vs.~$i$ or $z$), would need to be determined via other calibration methods.\footnote{Absolute photometry, i.e.~an overall
absolute flux scale --- while less critical for most present astrophysical results (for example, measurements of dark energy using SNeIa
are largely independent of overall absolute flux scale) ---
would also not be addressed with an LPRS, and thus would also need to be determined using other methods.}
Stellar standards provide the typical means~\citep[e.g.][]{Boh04, Hol06}
for such other calibrations (in addition to individual laboratory characterization of filter, detector, camera, and telescope optical
spectral response, as described in~\citet{Doi10}, \citet{StuTon06}, and other references).
Another technique for characterizing combined relative throughput of telescope, camera, and detector, but not of the atmosphere, at pairs
of wavelengths across the optical spectrum is described in~\citet{Stu10}; and general techniques presently under development for precisely
measuring total throughput, and performing both relative and absolute spectrophotometry,
are described in~\citet{Albert12} and in~\citet{Alb21b}.  Conceivably, the upper-atmospheric Rayleigh backscattering
of light launched into the sky from a broadly-tunable laser --- and continuously monitored via a calibrated photodiode that receives a constant
small fraction of the laser light from a beamsplitter at this laser source --- could also be used for characterizing combined relative throughput;
although such a technique has, to our knowledge,
never been published.  The LPRS techniques we describe in this pair of papers are, thus, not the only techniques one would need to satisfy
all spectrophotometric calibration requirements; however they would provide a very high-precision reference to pinpoint the total relative
throughput at 589/590~nm vs.~at 819/820~nm.

\subsubsection{Laser pulse repetition rate; and pulse chirping} The pump laser for the LPRS will be pulsed at
a high frequency, such as the repetition rate of $\sim$250~MHz for the IPG~VLR-532-1000 mentioned earlier in this Section.  In order to
maximise the brightness of the LPRS~\citep{Kan14}, this repetition rate could be adjusted (and also re-adjusted as often as is practical) to be an integer
multiple of the Na atomic Larmor frequency
\begin{equation}
\qquad \qquad \qquad \qquad \quad f_{\rm Na,\:Larmor} \;\; \equiv \;\; \frac{\gamma_{\rm \;\!Na}}{2\pi} B,
\end{equation}
where the ground-state gyromagnetic ratio for sodium is $\frac{\gamma_{\rm \;\!Na}}{2\pi} = 699.812$~kHz/gauss.
For typical geomagnetic field strengths of $B = 0.25$~gauss to 0.5~gauss,
$f_{\rm Na,\:Larmor}$ can range between 175~kHz and 350~kHz (with a typical value being $f_{\rm Na,\:Larmor} \approx 260$~kHz),
and thus it would not be difficult to adjust the very high ($\sim$250~MHz) and continuously-adjustible repetition rate of the pump
laser to be an integer multiple of $f_{\rm Na,\:Larmor}$.  Either alternatively to, or in combination with, this adjustment of
the pump laser repetition rate, the laser output can have circular polarization orientation ($\sigma_+$ or $\sigma_-$)
that is modulated at $f_{\rm Na,\:Larmor}$ or an integer multiple thereof, in order to achieve a similar LPRS brightness-increasing
effect~\citep{Fan16,Ped18}.  Although, in contrast to typical LGS, only a very small fraction of the \ion{Na}{i} atoms in the sodium layer
will be excited by the laser at any given time in this LPRS, it would still be at least slightly beneficial to use these Larmor frequency  
techniques to close the excitation/de-excitation cycle between a specific pair of hyperfine states for a given \ion{Na}{i} atom, for the small
fraction of \ion{Na}{i} atoms in this LPRS that happen to repeat the cycle.
Another technique that also has the potential to at least marginally increase LPRS brightness is the chirping of the frequency of
LPRS pulses, similar to LGS continuous wave (CW) laser chirping as described in~\citet{Ped20}.  Although this LPRS laser will be a rapidly-pulsed
quasi-continuous-wave (QCW) source, rather than a CW source, the technique described in~\citet{Ped20} may be also applicable to a QCW LPRS by  
using frequency chirping either within each single pulse, or having the frequency increase during each chirp continue during a train
of multiple pulses.  However, in our estimations of observed flux and of impact on measurements of dark energy in the
following Sections, we do not assume that any of the possible LPRS brightness-increasing techniques described in this paragraph
have been implemented.

\hypertarget{sec6}{
\section{Estimation of Observed Signal Flux from LPRS}
}
\label{sec:EstFlux}

In this Section we calculate the expected observed flux at an observatory, for example the Rubin Observatory, which is located at the same mountaintop
site as a source laser with properties described in the previous Section, from the resulting de-excitation light generated in the
sodium layer in the mesosphere.  For a laser with 500~W average output power that is tuned to the $\lambda_s \equiv 342.78$~nm \ion{Na}{i} excitation,
the number of photons produced per second by the laser will be
$N^{\rm{laser}}_{\gamma} \equiv
\frac{P \lambda_s}{hc} = \frac{(500\:\rm{W}) \times (3.4278 \times 10^{-7}\:\rm{m})}{(6.63 \times 10^{-34}\:\rm{J\,s}) \times (3.0 \times 10^{8}\:\rm{m\,s}^{-1})}
= 8.61 \times 10^{20}$~photons/s.
The atmospheric transmission above the 2663~m Cerro Pachon site (the Rubin Observatory site, as our example) at 342.78~nm is approximately 60\% (with losses dominated 
by Rayleigh scattering) in a vertical path, thus the number of 342.78~nm photons that reach the upper atmosphere per second will be approximately
$N^{\rm{laser\mbox{-}to\mbox{-}mesosphere}}_{\gamma} \equiv 0.60 \times N^{\rm{laser}}_{\gamma} = 5.17 \times 10^{20}$~photons/s.  Of these
$N^{\rm{laser\mbox{-}to\mbox{-}mesosphere}}_{\gamma} = 5.17 \times 10^{20}$ photons each second, a small fraction will be absorbed
by \ion{Na}{i} atoms in the mesosphere; we determine this fraction in the following.  As was mentioned in Section~\ref{sec:ExtInfAndNaLayer}, the total column
density of neutral sodium atoms varies within a range of approximately $2 \times 10^{9}$~atoms/cm$^2$ to $8 \times 10^{9}$~atoms/cm$^2$; we shall use
$4 \times 10^{9}$~atoms/cm$^2 = 4 \times 10^{13}$~atoms/m$^2$ in our calculation.  The total absorption cross-section per ground state \ion{Na}{i} atom
$\sigma_{342.78\:\rm{nm}} = \frac{c^2}{8 \pi \nu_{ki}^2} \frac{g_k}{g_i} \phi_{\nu} A_{ki},$ where $A_{ki} = 2.1 \times 10^{2}$~s$^{-1}$ is the
Einstein $A$ coefficient,\textsuperscript{\ref{foot:EinA}}
$\nu_{ki} = \frac{c}{342.78\:\rm{nm}} = 1.14 \times 10^{15}$~s$^{-1}$, the level degeneracy ratio
$\frac{g_k}{g_i} = \frac{4}{2} = 2$, and $\phi_{\nu} \approx \frac{1}{1.5\:\rm{GHz}} = 6.67 \times 10^{-10}$~s.  Thus
$\sigma_{342.78\:\rm{nm}} \approx 7.74 \times 10^{-22} \frac{{\rm m}^2}{\rm atom}$, and thus each 342.78~nm photon has an approximately
($7.74 \times 10^{-22} \frac{{\rm m}^2}{\rm atom}) \times (4 \times 10^{13} \frac{\rm atoms}{{\rm m}^2}) = 3.10 \times 10^{-8}$ chance of exciting an \ion{Na}{i} atom.  
There will thus be approximately $3.10 \times 10^{-8} \times N^{\rm{laser\mbox{-}to\mbox{-}mesosphere}}_{\gamma} =$
\begin{equation}
\qquad \qquad \quad \; 1.6 \times 10^{13}\;{\rm \ion{Na}{i} \; atoms \; excited \; per \; second}
\end{equation}
in the mesosphere.

Each of those excited \ion{Na}{i} atoms will emit one 818.55~nm or 819.70~nm photon, as well as one 589.16~nm or 589.76~nm photon, within approximately 100~ns,
with the photons emitted in uniform angular distributions.
The atmospheric transmission back down to the 2663~m Cerro Pachon site at wavelengths of $\sim$819~nm and $\sim$589.5~nm is approximately 90\% in the
case of both of those wavelengths (with losses dominated by water vapor absorption and by Rayleigh scattering respectively), and thus at
the telescope approximately
95~km below the sodium layer, this will correspond to approximately
$N^{\rm{signal}}_{\gamma} \equiv 0.9 \times (1.6 \times 10^{13} \frac{\rm photons}{\rm s}) \times \frac{1}{4\pi \times (9.5 \times 10^{4}\,{\rm m})^{2}} =$
\begin{equation}
\qquad \qquad \qquad \qquad \quad 127\;{\rm photons/s/m}^2
\end{equation}
at each of 818.55 or 819.70~nm, and 589.16 or 589.76~nm.

The apparent magnitude $m \equiv -2.5 \log_{10}\!\left(\frac{\phi}{\phi_0}\right)$, and in the AB reference system
$m_{\rm AB} \equiv -2.5\log_{10}(\phi) - 48.6$ when the spectral flux density $\phi$ is provided in units of ${\rm erg\:s^{-1}\:cm^{-2}\:Hz^{-1}}$.
The bandwidths $b$ of the Rubin Observatory $r$ and $z$ filters are approximately (549~--~694)~nm and (815~--~926)~nm respectively, which
correspond to approximate bandwidths of $b^{r\:{\rm band}} = 1.14 \times 10^{14}$~Hz and $b^{z\:{\rm band}} = 4.41 \times 10^{13}$~Hz respectively.
These filter bandwidths must be multiplied by filter spectral ratios
$f \equiv$~(average filter throughput over the band)/(filter throughput at the wavelength of the nearly-monochromatic light)
to account for the 818.55/819.70~nm light being near the edge of the $z$ band, whereas the 589.16/589.76~nm light is near the centre of the $r$ band;
$f^{r\:{\rm band}}_{589.5\:{\rm nm}} = \frac{0.881}{0.943} = 0.934$ and $f^{z\:{\rm band}}_{819\:{\rm nm}} = \frac{0.851}{0.501} = 1.699$, and thus the
corrected effective bandwidths are
$f^{r\:{\rm band}}_{589.5\:{\rm nm}} \times b^{r\:{\rm band}} = 1.06 \times 10^{14}$~Hz and
$f^{z\:{\rm band}}_{819\:{\rm nm}}   \times b^{z\:{\rm band}} = 7.49 \times 10^{13}$~Hz respectively.
At a wavelength of 589.5~nm, 127~photons/s/m$^2$ corresponds to $4.29 \times 10^{-14}\:{\rm erg\:s^{-1}\:cm^{-2}}$, and at a wavelength of 819~nm,
127~photons/s/m$^2$ corresponds to $1.55 \times 10^{-14}\:{\rm erg\:s^{-1}\:cm^{-2}}$.  Thus, for an LPRS, 
$\phi^{r\:{\rm band}} = (4.29 \times 10^{-14}\:{\rm erg\:s^{-1}\:cm^{-2}})/(1.06 \times 10^{14}\:{\rm Hz}) = 4.05 \times 10^{-28}\:{\rm erg\:s^{-1}\:cm^{-2}\:Hz^{-1}}$ 
and
$\phi^{z\:{\rm band}} = (3.08 \times 10^{-14}\:{\rm erg\:s^{-1}\:cm^{-2}})/(7.49 \times 10^{13}\:{\rm Hz}) = 4.11 \times 10^{-28}\:{\rm erg\:s^{-1}\:cm^{-2}\:Hz^{-1}}$; 
and thus
\begin{eqnarray}
m^{r\:{\rm band}}_{\rm AB} & = & 19.9,\;{\rm and}\\[2mm]
m^{z\:{\rm band}}_{\rm AB} & = & 19.9
\end{eqnarray}
for an LPRS.  Since 819~nm light lies in the overlap region between the LSST
$i$ and $z$ filters, that light from the LPRS will additionally be (more dimly) observable 
in the LSST $i$ band.  An analogous calculation to the above thus gives
$\phi^{i\:{\rm band}} = (3.08 \times 10^{-14}\:{\rm erg\:s^{-1}\:cm^{-2}})/(1.43 \times 10^{14}\:{\rm Hz}) = 2.15 \times 10^{-28}\:{\rm erg\:s^{-1}\:cm^{-2}\:Hz^{-1}}$ 
and
\begin{equation}
\qquad \qquad \qquad \qquad \quad m^{i\:{\rm band}}_{\rm AB} \quad = \quad 20.6\phantom{,\;{\rm and}}
\end{equation}
for an LPRS.

As determined above, the flux of LPRS signal photons incident on the telescope is 127~photons/s/m$^2$ at each of
819~nm and 589~nm, which would correspond to 4445~photons/s at each of those two wavelengths in the specific case of the 35~m$^2$ clear aperture of the
Simonyi Survey Telescope at the Rubin Observatory.  Using the expected LSST values for instrumental optical throughput and detector quantum efficiency~\citep{Jones19}, this 
corresponds to total numbers of observed signal photoelectrons equal to $5.73 \times 10^4$, $2.40 \times 10^4$, and $3.73 \times 10^4$ within the
elliptical LPRS spot (with the previously-calculated major and minor axis angular diameters of $2.1\arcsec \times 1.4\arcsec$)
during a 30~s visit, in the LSST $r$, $i$, and $z$ filters respectively. That corresponds to averages of 959, 402, and 625
signal photoelectrons per
$0.2\arcsec \times 0.2\arcsec$ LSST camera pixel at the centre of the Gaussian spot in that time interval;
i.e.~signal standard deviations at the centre of the Gaussian spot of approximately
$\sqrt{959} = 31.0$, $\sqrt{402} = 20.0$, and $\sqrt{625} = 25.0$
photoelectrons per pixel per visit in the LSST $r$, $i$, and $z$ filters respectively.

\begin{figure*}
\begin{center}
\includegraphics[scale=.332]{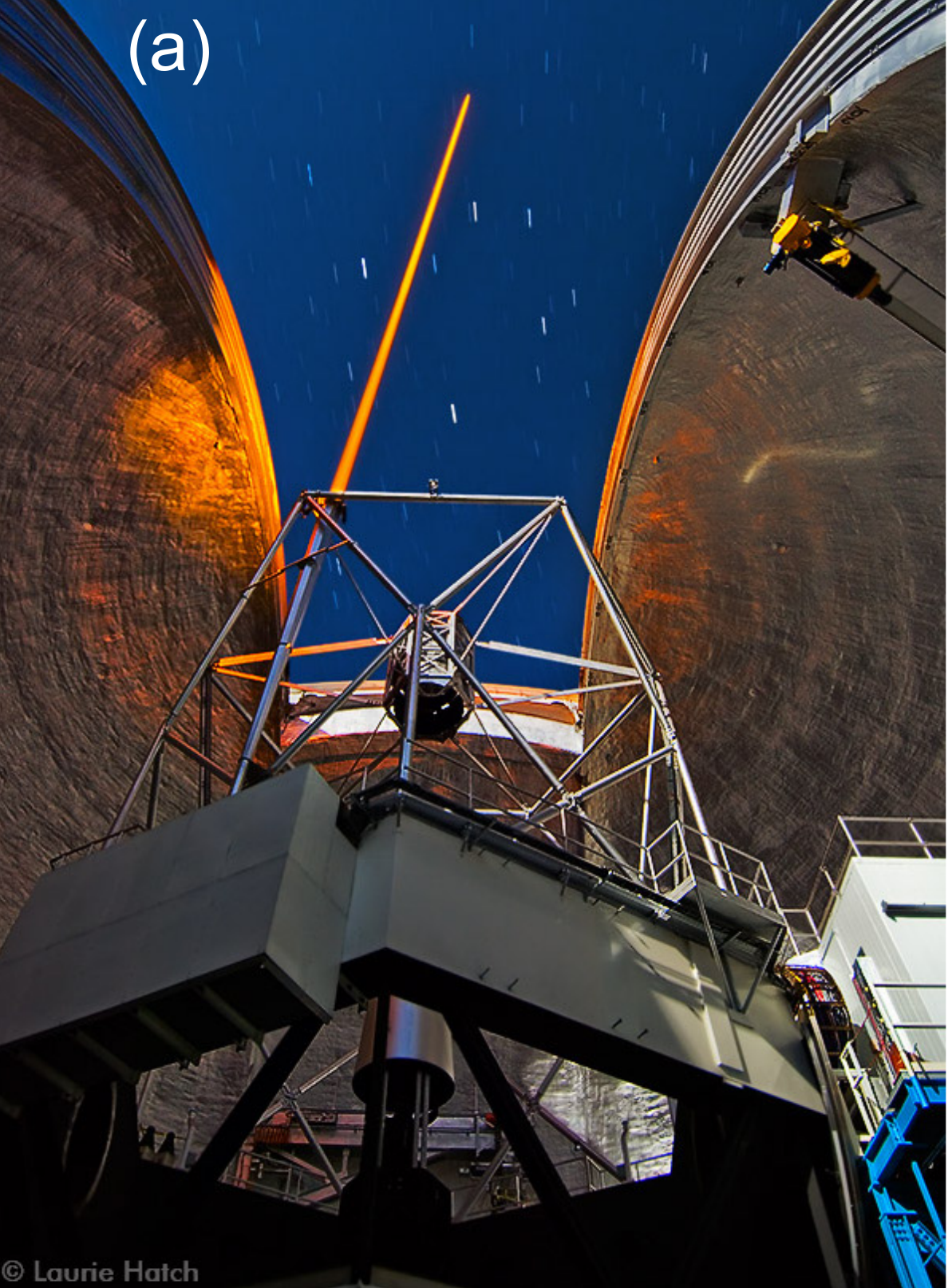}
\hspace{4mm}
\includegraphics[scale=.332]{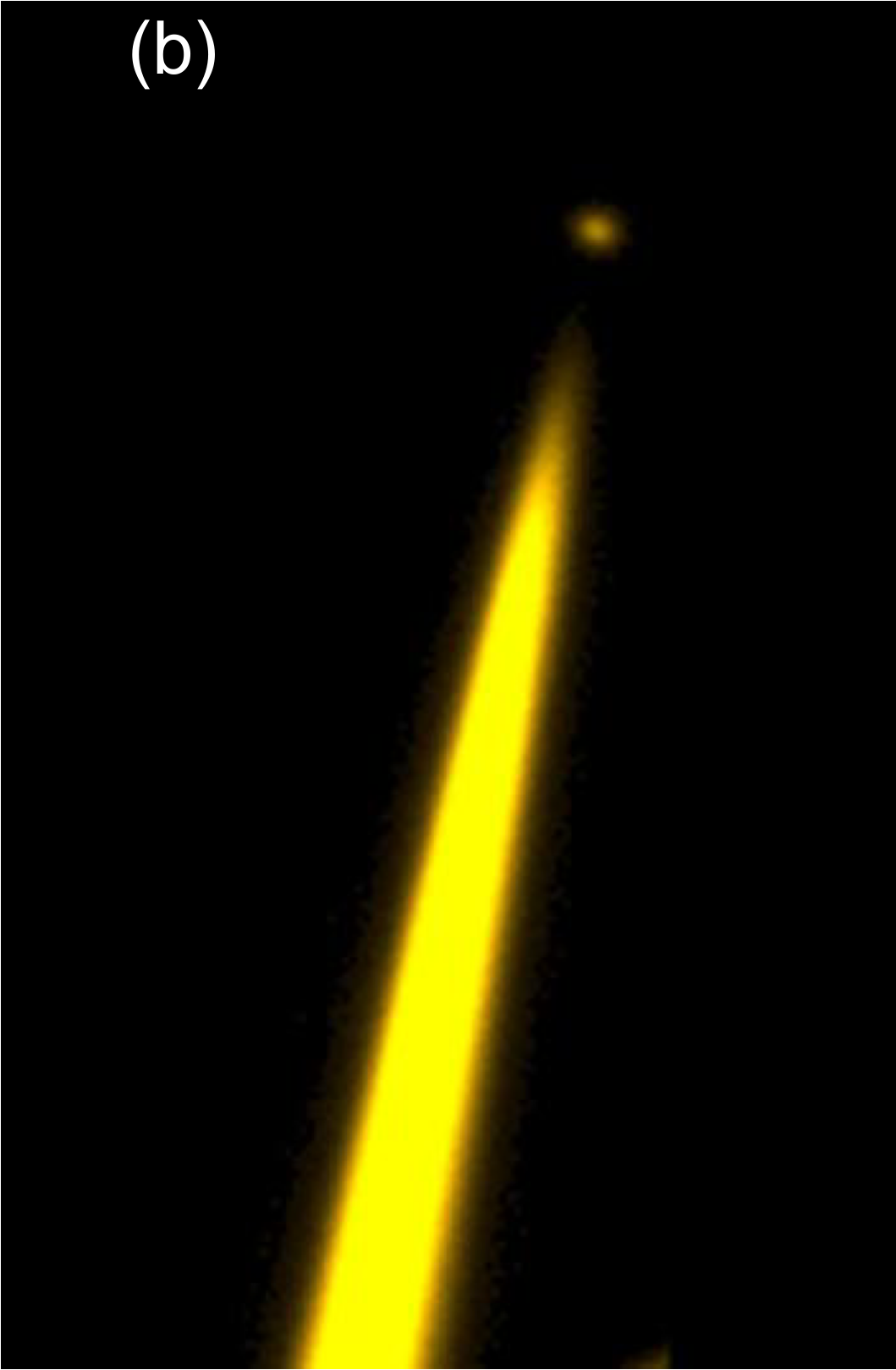}
\end{center}  
\caption{For visualization of the effects of Rayleigh scattering for LGS as well as for LPRS: (a)~A photograph of the Keck II LGS physical setup
(copyright Laurie Hatch, \url{https://lauriehatch.com}) during operation, and (b)~A pseudocolour image of an LGS beacon located near the
William Herschel Telescope on La Palma, Spain, taken using a small portable telescope placed 3~m from the laser launch site~\citep{LaPalmaLGS}.
The qualitative aspects of Rayleigh-scattered laser light for an LPRS would, for the most part, be similar to those for these and other
sodium LGS, however there are two differences that would tend to make the effects from Rayleigh scattering larger in an LPRS than in an LGS, and a third
that would tend to make Rayleigh scattering effects smaller in an LPRS than in an LGS: 1)~The relative amount of laser light, to signal light returned
from the sodium layer, is 5 orders of magnitude larger for LPRS than for LGS; 2)~The cross-section for Rayleigh scattering at 343~nm is approximately 9
times greater than at 589~nm; and 3)~Rayleigh-scattered 343~nm light is outside of the filter bandpasses of the two LPRS calibration lines near 589~nm
and 819~nm respectively.}
\label{fig:KeckRayleigh}
\end{figure*}

In any forseeable application of this technique, the dominant uncertainty on the relative amount of light from 818.55/819.70~nm vs.~589.16/589.76~nm signal photons
would be the Poisson uncertainties on small-sample collected photon statistics,\footnote{also known as shot noise} and the analogous binomial uncertainty on their ratio.
However, given hypothetical infinite photon statistics, one might ask just how well the 1:1 ratio is predicted, i.e.~what the dominant systematic uncertainty
on that signal ratio would be.  We estimate that the dominant systematic uncertainty on the 1:1 signal ratio would, by far, be due to collisions of the neutral
sodium atoms in the upper atmosphere during de-excitation: inelastic collisions of the excited atoms could potentially eliminate the production of (or dramatically modify the
wavelength of) either the 818.55/819.70~nm or the 589.16/589.76~nm photon during a given de-excitation.  The typical frequency of atomic collisions of a given sodium
atom at the $\sim$100~km altitude of the sodium layer is   
\begin{equation}
\label{eq:CollisionRate}
\qquad \qquad \nu_{\rm collision} \; = \; 4 n_{\rm 100\:km} \langle\sigma\rangle \sqrt{\frac{k_B T_{\rm 100\:km}}{\pi m_{\rm Na}}} \; \approx \; 770\;{\rm s}^{-1},
\end{equation}
where $n_{\rm 100\:km} \approx 10^{19}\:$m$^{-3}$ is the total number density of all species at 100~km altitude,
$\langle\sigma\rangle \approx$\linebreak
$\pi \times (2 \times 10^{-10}\:{\rm m})^2 = 4\pi \times 10^{-20}$~m$^2$ is an averaged cross-sectional area for the collisional interaction of an \ion{Na}{i}
atom with any other species at that altitude, and $T_{\rm 100\:km} \approx 200$~K.  The Einstein $A$ coefficients
of the 818.55~nm, 819.70~nm, 589.16~nm, and 589.76~nm transitions range from $A_{\rm min} = 8.57 \times 10^6$~s$^{-1}$ to
$A_{\rm max} = 6.16 \times 10^7$~s$^{-1}$, so we can set an upper bound on the fractional expected systematic deviation from a 1:1 ratio of 818.55/819.70~nm
vs.~589.16/589.76~nm signal photons as
\begin{equation}
\qquad \qquad \qquad \quad \:\:\: \epsilon \:\: \equiv \:\: \frac{\nu_{\rm collision}}{A_{\rm min}} \:\: = \:\: 9 \times 10^{-5},
\end{equation}
so that the ratio together with its systematic uncertainty becomes ($1 \pm \epsilon$):1.
(This bound could conceivably be made yet tighter via a detailed simulation of collisional processes.)

\hypertarget{sec7}{
\section{Estimation of LPRS Observed Background}
}
\label{sec:EstBkgd}

Observed light from an LPRS will be superimposed on two types of background light: $(a)$~Laser-induced background light; and $(b)$~The typical diffuse sky background
(which we, as is usual, take to include all other [i.e.~non-laser] light that is scattered by the optical elements of the telescope, as well as by the
atmosphere and by zodiacal dust), plus instrumental background noise.  In this Section we estimate the size of contributions from both of
these sources of background: in particular we show that type~$(a)$ may indeed be quite significant unless (342.6~--~342.8)~nm light can be rejected
by the $r$, $i$, and $z$ filters at a level (optical density [OD] of $>6.7$) that is significantly greater than the present LSST filter specifications, and that
type~$(b),$ the background over which all objects are observed, will be corrected through the usual technique of sky subtraction.

\begin{figure*}
\begin{center}
\includegraphics[scale=.26]{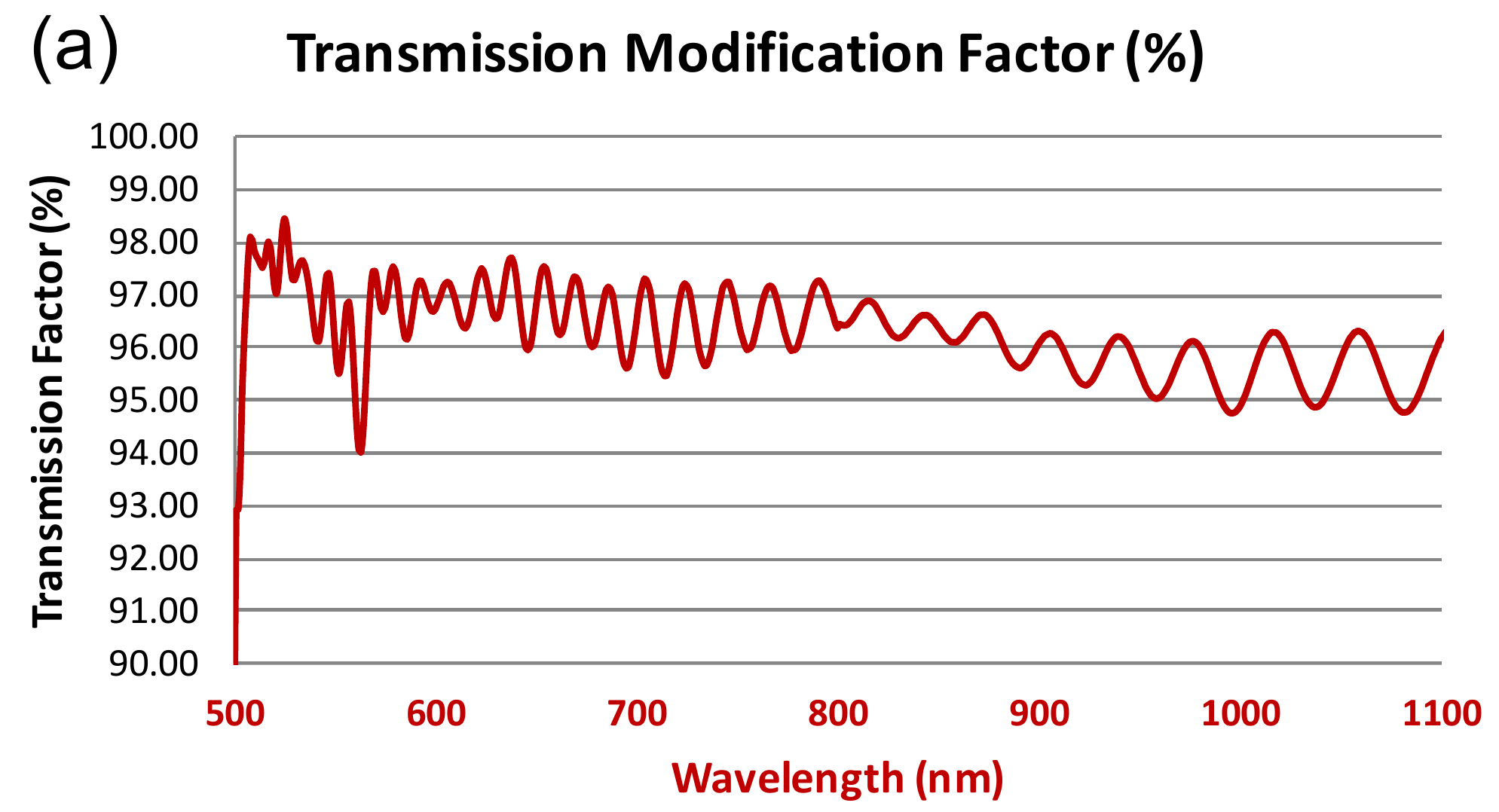}
\hspace{2mm}
\includegraphics[scale=.26]{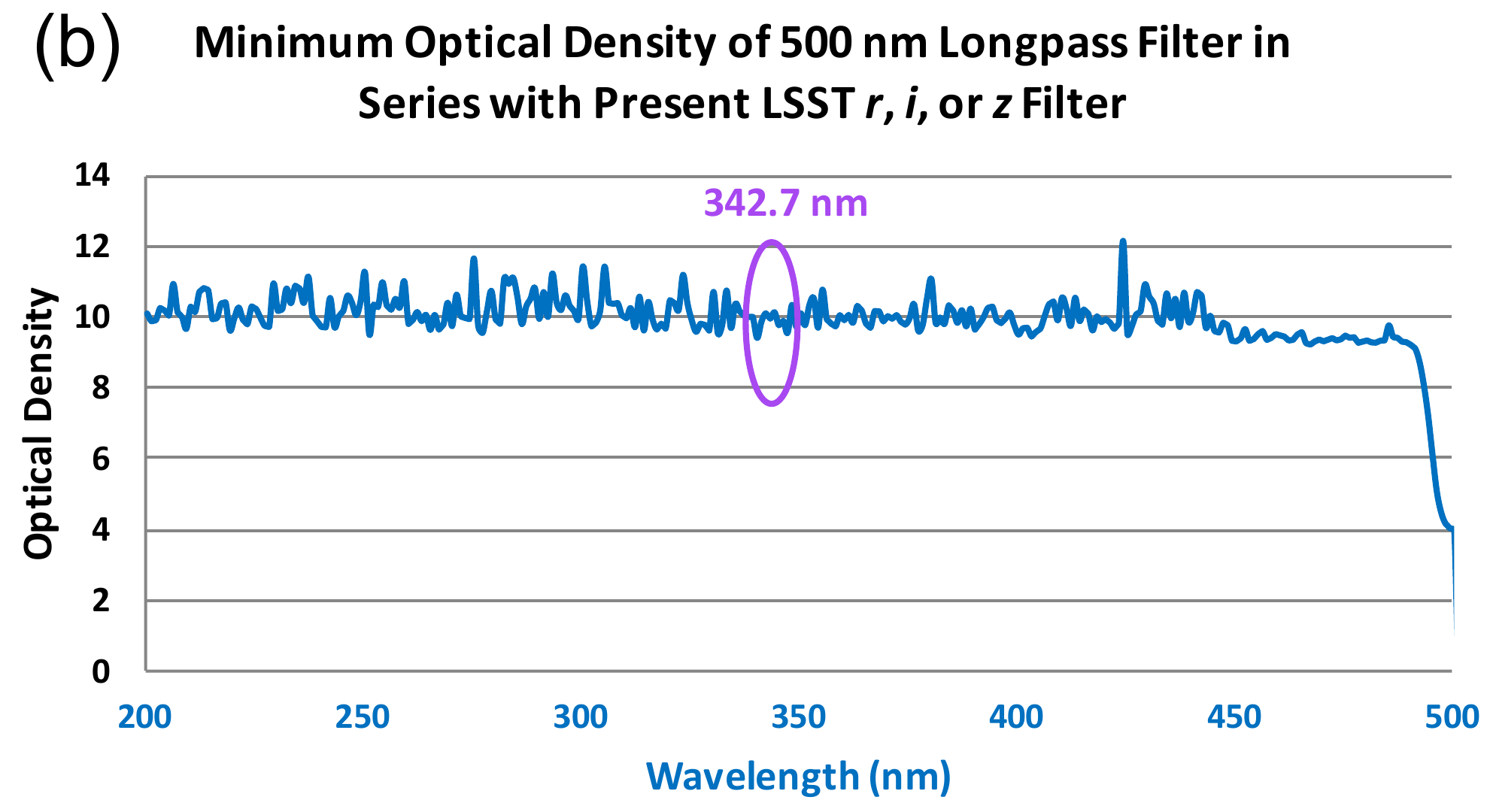}
\end{center}
\caption{(a)~The percentage of transmission that is preserved (relative to present LSST filters), as a function of wavelength
from 500~nm to 1100~nm; and (b)~The minimum optical density (OD) when out of bandpass, as a function of wavelength from 200~nm
to 500~nm, for a longpass filter with cutoff wavelength of 500~nm, when in series with the present LSST $r$, $i$, or $z$ filter.
The present LSST filters have specifications of minimum OD of 4 when out of bandpass; the addition of such a longpass filter
in series with the LSST $r$, $i$, or $z$ filter would raise the OD to greater than 10 for (342.6~--~342.8)~nm, while preserving
more than 94\% of present light at all in-bandpass wavelengths for each of those 3 filters.  This longpass filter data
from~\citet{Thorlabs} (product FELH0500).
}
\label{fig:FilterTrans}
\end{figure*}

For type~$(a),$ potential laser-induced background light can itself be divided into two categories:
$(a_1\!)$ The background light from near-180\degr~atmospheric Rayleigh scattering of
the 342.78~nm laser light that also manages to pass through the $r$, $i$, or $z$ filters (via an imperfect rejection of 342.78~nm light by those filters); and $(a_2\!)$ The
background light from near-180\degr~atmospheric Raman scattering and
de-excitation light from other inelastic excitations of atmospheric atoms and molecules in the laser's path, within the passbands of the $r$, $i$, or $z$ filters.
As we will calculate below, near-180\degr~Rayleigh scattering will result in a large flux of 342.78~nm photons into the telescope aperture, and thus
to reduce background of type~$(a_1\!)$ to a negligible level,
this technique would require the thorough rejection of 342.78~nm photons (at OD $>6.7$) by the $r$, $i$, and $z$ filters.

The product of the Rayleigh backscattering cross-section $\sigma^R_b$ with atmospheric molecular density $n(z)$ can be approximated by the convenient formula:
\begin{equation}
\label{eq:RayleighTot}
\qquad \quad \sigma^R_b n(z) \; = \; (3.6 \times 10^{-31})\frac{p(z)}{T(z)}\lambda^{-4.0117}\;{\rm m}^{-1}{\rm sr}^{-1},
\end{equation}
where $p(z)$ is the pressure in millibars at an altitude $z$ above sea level, $T(z)$ is the temperature in Kelvin at altitude $z$, and
both $z$ and $\lambda$ are in units of metres~\citep{Cer80}.  Using a roughly approximate isothermal atmosphere model
$\frac{p(z)}{T(z)} \approx \frac{p_0}{T_0} e^{-\left( \frac{\mu g z}{R T_0} \right)}
\approx \frac{1000}{300}e^{-\left( \frac{z}{8800} \right)}$,
we can integrate equation~(\ref{eq:RayleighTot}) over both the altitude above the 2663~m telescope site and by the solid angle that is
subtended by the telescope aperture to obtain an approximate fraction of laser photons that are Rayleigh-backscattered
by the atmosphere into the aperture of the telescope:
\begin{eqnarray}
f_{b, {\rm total\:streak}}^{R,\gamma}
                         & \approx & \frac{\scriptstyle 20\pi \times (3.6 \times 10^{-31}) \times (3.428 \times 10^{-7})^{-4.0117}}{\scriptstyle 3} \times \nonumber\\
                         &         & \int^\infty_{z = 2663} e^{-\left( \frac{z}{8800} \right)} dz
                                     \int^{\tan^{-1}(\frac{r}{z - 2663})}_{\theta = 0}
                                       \sin\theta d\theta \; \qquad                                        \\
                         & \approx & 2 \times 10^{-3},                                                     \nonumber
\end{eqnarray}
where $r = 4$~m is the radius of the telescope aperture.  That would naively indicate that the $r$, $i$, and $z$ filters would need to filter out
at the very least $1 - \frac{N^{\rm{signal}}_{\gamma}}{f_{b, {\rm total\:streak}}^{R,\gamma} N^{\rm{laser}}_{\gamma}} \approx 1 - (7.4 \times 10^{-17})$
of all such photons, i.e.~have OD of
$-\log_{10}(7.4 \times 10^{-17}) = 16.1$ for $\lambda \in$~(342.6 -- 342.8)~nm.
However, since, as shown in Fig.~\ref{fig:KeckRayleigh}(a), the centre of the laser launch telescope would be slightly offset from the centre of
the aperture of the observing telescope, the large majority of those Rayleigh-backscattered laser photons that enter the telescope aperture, even
if they also manage to get through the $r$, $i$, or $z$ filter, would not be directly superimposed on the signal photons returned from the sodium-layer
LPRS; the Rayleigh-backscattered light instead tends to form a streak that ``leads up'' to the light that is returned from the sodium layer, as shown for
example in Fig.~\ref{fig:KeckRayleigh}(b), but is not actually superimposed on the LPRS spot.  Thus, the only such Rayleigh-backscattered background photons that
would, in fact, be superimposed on the photons from the LPRS on the focal plane (if such photons happened to get through the $r$, $i$, or $z$ filter) would
be photons that are Rayleigh-backscattered from the same altitude range as the sodium layer itself, i.e.~from between 80 and 105 km above sea 
\hypertarget{eq12}{level:}
\begin{eqnarray}
f_{b, {\rm over\:LPRS\:spot}}^{R,\gamma}  \!\!\!\!
                         & \approx & \frac{\scriptstyle 20\pi \times (3.6 \times 10^{-31}) \times (3.428 \times 10^{-7})^{-4.0117}}{\scriptstyle 3} \times \nonumber\\
                         &         & \int^{105\,000}_{z = 80\,000} \!\! e^{-\left( \frac{z}{8800} \right)} dz
                                     \int^{\tan^{-1}(\frac{r}{z - 2663})}_{\theta = 0}
                                       \!\!\!\! \sin\theta d\theta \quad \qquad                             \\
                         & \approx & 7 \times 10^{-13},                                                     \nonumber
\end{eqnarray}
and thus the $r$, $i$, and $z$ filters only must filter out at very least
$1 - \frac{N^{\rm{signal}}_{\gamma}}{f_{b, {\rm over\:LPRS\:spot}}^{R,\gamma} N^{\rm{laser}}_{\gamma}} \approx 1 - (2.1 \times 10^{-7})$
of all such photons, i.e.~have OD of at very least $-\log_{10}(2.1 \times 10^{-7}) = 6.7$ for $\lambda \in$~(342.6 -- 342.8)~nm.
This requirement, while still stringent, is vastly less stringent than a requirement of OD of more than 16.
However, the baseline LSST filter specification only requires that the LSST filters reject of out-of-band light at OD $>4$~\citep{LSSTOverview}
and, thus, achieving a notch rejection of OD $>6.7$ for $\lambda \in$~(342.6~--~342.8)~nm light would likely require a future upgrade of
those three LSST filters.  Such an upgrade of those filters would not be technically difficult to achieve; for example a standard longpass
filter [such as, e.g., product FELH0500 from~\citet{Thorlabs}]
to remove wavelengths below 500~nm in series with the present $r$, $i$, and $z$ filters would preserve more than 94\% of all present
pass-through light across all within-bandpass wavelengths in all 3 filters, while simultaneously achieving a total OD of greater than 10 for
(342.6~--~342.8)~nm light, as shown in Fig.~\ref{fig:FilterTrans}.
Single combination longpass $r$, $i$, and $z$ wavelength filters
could likely be produced with the same physical thicknesses and bandpasses as those present LSST filters, but also combining the strong
rejection of $\lambda < 500$~nm light of the longpass
filter.\footnote{Alternatively, polarization filters, rather than (or even in addition to) longpass wavelength filters, could be used to
greatly reduce laser atmospheric Rayleigh-backscattered light.  However, such filters would necessarily only admit light from one
polarization from astronomical sources, as well as from the LPRS, which may be unwanted for some astronomical observations.  
\citetalias{Alb21a} considers an alternative LPRS generation scheme, which (unlike the scheme in this paper)
would necessarily require polarizing filters to be installed, however would provide a brighter LPRS than the technique
described in this paper.}

Laser-induced background of type $(a_2\!)$, i.e.~near-180\degr~atmospheric Raman scattering and other inelastic collisions of 342.78~nm laser light
into the passbands of the $r$, $i$, or $z$ filters,
would systematically affect LPRS-based photometric calibration.
In particular, molecular oxygen has a column density which is more than 10 orders of magnitude larger than the \ion{Na}{i} column density in the sodium layer.
Photons scattered from O$_2$ with wavelengths of 762~nm and 688~nm arise from spin-forbidden transitions, and while Raman lines in the Schumann-Runge
bands are strong, the above lines produce cross-sections of approximately 10$^{-40}$~cm$^2$ per molecule, and can therefore safely be ignored.

The recent observations, using the 4 Laser Guide Star Facility
(4LGSF)~\citep{Cal14}, of O$_2$ and N$_2$ Raman rotational transitions
corresponding to the first vibrational excitation in these
molecules~\citep{Vog17}, reveal that at the maximum of resonant Raman
signal for the $\Delta J = 0$ transitions, the line intensity is approximately $10^{-16}$~erg~s$^{-1}$~cm$^{-2}$.
Line intensities due to off-resonant scattering from light at 343~nm will be several orders of magnitude smaller than that value, and thus will be
negligible when compared with the returned signal flux from the sodium layer (which has an intensity at 589.16/589.76~nm
calculated in Section~\ref{sec:EstFlux} of approximately $4.29 \times 10^{-14}$~erg~s$^{-1}$~cm$^{-2}$).

The typical non-laser-induced diffuse sky background, i.e.~background of type~$(b)$, would be corrected via standard techniques of sky subtraction.
As was shown in Section~\ref{sec:Laser}, the elliptical LPRS spot would have a major axis of approximately 2.1\arcsec~angular diameter and minor
axis of approximately 1.4\arcsec~angular diameter on the sky, which is only slightly larger than the PSF of the LSST camera at the Rubin Observatory
when including effects of atmospheric
seeing, which is approximately 1\arcsec~angular diameter in total~\citep{LSSTOverview}. The sky brightness at the zenith in the case of the Rubin Observatory 
site is estimated to be, in magnitude per square arcsecond, 21.2, 20.5, and 19.6 for the $r$, $i$, and $z$ filters respectively~\citep{Jones17}, which corresponds
to approximately 1498, 2151, and 3544 photons/s/(square~arcsecond) for the three filters respectively, thus
3459, 5141, and 8472 photons/s within an elliptical spot of angular diameter $2.1\arcsec \times 1.4\arcsec$.

When using the expected LSST values for instrumental optical throughput and detector quantum efficiency as
functions of wavelength~\citep{Jones19}, those values of 3459, 5141, and 8472 photons/s correspond to total numbers of observed sky background photoelectrons
equal to $4.46 \times 10^4$, $6.63 \times 10^4$, and $1.07 \times 10^5$ within the elliptical spot during a 30~s visit consisting of two 15~s exposures,
in the $r$, $i$, and $z$ filters respectively.
This corresponds to 747, 1110, and 1701 photoelectrons per 0.2\arcsec~$\times$~0.2\arcsec~pixel in that time interval; i.e.~sky background
standard deviations of approximately $\sqrt{747} = 27.3$, $\sqrt{1110} = 33.3$, and $\sqrt{1701} = 41.2$ photoelectrons per pixel per visit in
the $r$, $i$, and $z$ filters respectively.  In the same 30~s visit time interval, the expected standard deviation in each pixel
due to instrumental background noise (in each of the filters of course) is 12.7 photoelectrons~\citep{Jones17}.

\section{Resulting Estimated Photometric Ratio Precision}
\label{sec:ResPrec}

\begin{figure*}
\begin{center}
\includegraphics[scale=.65]{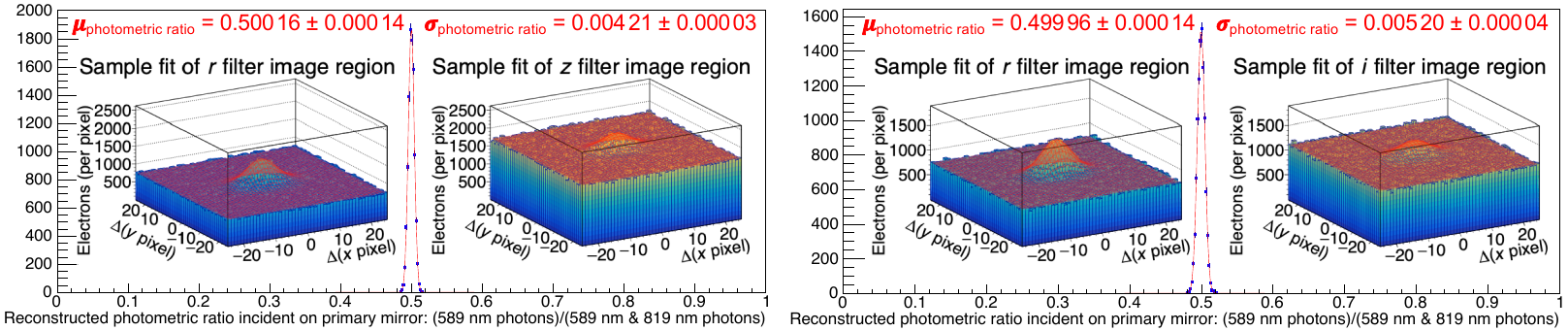}
\end{center}
\caption{
Within both the left and right sets of plots above, the main plot (i.e., the 1-dimensional fitted Gaussian curve in the centre of each set of plots) 
shows the distribution of $10\,000$ fitted photometric ratios, with each ratio reconstructed from
pairs of simulated $r$ and $z$ filter image regions (within the set of plots on the left-hand side above),
and pairs of simulated $r$ and $i$ filter image regions (within the set of plots on the right-hand side above).
For both the left and right sets of plots,
each one of the $4 \times 10\,000$ simulated image regions (in total) consists of a 50~pixel~$\times$~50~pixel square centered
around the observed LPRS centroid.  The inset plots show single
examples of simulated $r$ and $z$ filter image regions (within the set of plots on the left),
and simulated $r$ and $i$ filter image regions (within the set of plots on the right),
and their respective fits to 2-dimensional Gaussian ellipses plus flat background
distributions.  The photoelectrons
and noise electrons in each simulated pixel, in each region, are generated according to the statistical distributions expected from a single 30~s
LSST visit consisting of a pair of 15~s exposures.  The fitted number of signal photoelectrons is extracted from each image region fit,
and divided by the (photoelectron)/(incident photon) efficiency (consisting of the expected detector quantum efficiency of the LSST camera at the Rubin Observatory, 
multiplied by the expected
throughput fraction of telescope, camera, and filter optics, at 589~nm and at 819~nm for the simulated $r$, $z$, and $i$ filter image regions respectively),
to determine the reconstructed numbers of 589~nm and 819~nm photons incident on the Rubin Observatory primary mirror during the 30~s visit.
For each simulated image region pair, the resulting ratio of reconstructed (589~nm photons)/(589~nm photons + 819~nm photons) is plotted,
and the resulting distribution is fitted to a single Gaussian.  The standard deviation of the fit
that is shown in the main plot within the \textit{left} set of plots above, which corresponds to the expected
LPRS photometric ratio statistical uncertainty from a single pair of visits with the $r$ and $z$ filter, is equal to $0.004\,21 \pm 0.000\,03$, and
the mean of the fit is consistent with 0.5.
The standard deviation of the fit that is
shown in the main plot within the  \textit{right} set of plots above, which corresponds to the expected
LPRS photometric ratio statistical uncertainty from a single pair of visits with the $r$ and $i$ filter, is equal to $0.005\,20 \pm 0.000\,04$, and
the mean of that fit is also consistent with 0.5.
}
\label{fig:rziphotratiosimfit}
\end{figure*}

We determine the resulting estimated photometric ratio measurement precision in the case of utilizing a pair of 30~s LPRS visits in the LSST $r$ and $z$ filters, and
also in the case of utilizing a pair of visits in the LSST $r$ and $i$ filters, in two ways: (1)~Using a simple analytic approximation via error propagation;
and (2)~Using a more detailed numerical determination, using sets of simple Monte Carlo simulations with the
ROOT software package~\citep{ROOT,ROOTnew}.  Both of these estimates assume that Rayleigh-scattered out-of-band 343~nm laser photons are fully blocked from entering the
$r$, $i$, or $z$ filter images, via longpass upgrades of those three filters, as described in the previous section.

We first make simple analytic approximations of the precision of measurements of the photometric ratio (589~nm photons)/(589~nm photons + 819~nm photons)
using pairs of 30~s visits to the LPRS spot, per~(1).  Let the reconstructed number of 589~nm photons incident on the primary mirror during a
30~s $r$ filter visit equal $\rho \pm \delta_{\!\rho}$, and the reconstructed number of 819~nm photons incident on the primary mirror during a
30~s $z$ filter visit equal $\zeta \pm \delta_{\!\zeta}$.
We are then measuring the ratio $\frac{\rho \pm \delta_{\!\rho}}{(\rho \pm \delta_{\!\rho}) +(\zeta \pm \delta_{\!\zeta})}$
which, via error propagation, is equal to $\frac{\rho}{\rho + \zeta} \pm \frac{\zeta\delta_{\!\rho} + \rho\delta_{\!\zeta}}{(\rho + \zeta)^2}$,
when $\frac{\delta_{\!\rho}}{\rho}$ and $\frac{\delta_{\!\zeta}}{\zeta}$ are both small and the uncertainties are both uncorrelated
and Gaussian distributed.  From the calculations in Section~\ref{sec:EstFlux}, $\rho$ and $\zeta$ will both be equal to approximately
$30 \times 4445 = 133\,350$ photons.  The uncertainties $\delta_{\!\rho}$ and $\delta_{\!\zeta}$ will approximately equal the square root of the total number of electrons
measured within the detector area of the LPRS spot during the respective visits, each scaled by the ratio of incident photons to measured signal photoelectrons
(i.e.~by the inverse of the telescope throughput fraction [including detector quantum efficiency]) at the respective wavelengths of 589~nm and 819~nm.  
Using the calculations in Sections~\ref{sec:EstFlux} and~\ref{sec:EstBkgd}, we can see that $\delta_{\!\rho}$ will be approximately
$\sqrt{57\,300 + 44\,600} \times \frac{133\,350}{57\,300} = 742.9$~photons and $\delta_{\!\zeta}$ will be approximately
$\sqrt{37\,300 + 107\,000} \times \frac{133\,350}{37\,300} = 1358.1$~photons,
and thus $\frac{\rho}{\rho + \zeta} \pm \frac{\zeta\delta_{\!\rho} + \rho\delta_{\!\zeta}}{(\rho + \zeta)^2} \approx 0.5 \pm 0.003\,94$, 
i.e.~a fractional uncertainty of just under a part in 100.  A similar calculation for a pair of visits in the $r$ and $i$ filters (rather than in the $r$ and $z$ filters) yields
$(\iota \pm \delta_{\!\iota}) \approx (133\,350 \pm 1670)$~photons, and thus
$\frac{\rho}{\rho + \iota} \pm \frac{\iota\delta_{\!\rho} + \rho\delta_{\!\iota}}{(\rho + \iota)^2} \approx 0.5 \pm 0.004\,52$, also an uncertainty of just under a
part in 100, and just slightly larger than when using a pair of visits in the $r$ and $z$ filters.
As we will see, these simple analytic approximations only slightly underestimate the statistical uncertainties, when they are compared with numerical simulations.

Figure~\ref{fig:rziphotratiosimfit} shows the results of sets of numerical simulations, per method~(2), to determine the precision of photometric
ratio measurement using single pairs of visits to the LPRS spot.  As shown in this figure, and in the description in its caption, the resulting estimates of the
photometric ratio measurement and its
statistical uncertainty are $0.5 \pm 0.004\,21$ (the analytic approximation above gave $0.5 \pm 0.003\,94$), and $0.5 \pm 0.005\,20$
(the analytic approximation above gave $0.5 \pm 0.004\,52$), for pairs of visits in the $r$ and $z$ filters, and in the $r$ and $i$ filters, respectively.

Those statistical uncertainties could, of course, be reduced by utilizing more than a single pair of visits, at the cost of increased observation time.  Increasing the
average optical output power of the LPRS laser above the nominal 500~W would also reduce the statistical uncertainties, by increasing the number of observed LPRS photons.

\hypertarget{sec9}{
\section{Estimated Impact on Measurements of Dark Energy from Type Ia Supernovae}
}
\label{sec:EstImp}

We estimate the impact of the improved photometry, when utilizing the
photometric ratios provided by this LPRS at a survey
observatory such as the Rubin Observatory, on future measurements
using type~Ia supernovae of the dark energy equation of state as a function of redshift $w(z)$.
The function $w(z)$ is defined as:
\begin{equation}
\qquad\qquad\qquad\qquad\quad w(z) \; \equiv \; p_{\rm DE}(z)/\rho_{\rm DE}(z),
\end{equation}
where $p_{\rm DE}(z)$ and $\rho_{\rm DE}(z)$ are respectively the pressure and the energy density of dark energy,
both in dimensionless units,
as functions of redshift $z$ (under the typical assumption that dark energy behaves as a perfect fluid).
In our simulations, we use the usual parametrization~\citep{Lin03}:
\begin{equation}
\qquad\qquad\qquad\qquad\quad w(z) \; = \; w_0 + \frac{z}{1 + z}w_a,
\end{equation}
where the quantities $w_0$ and $w_a$ respectively parameterise the equation of state of dark energy at the present time, and the
amount of change in the equation of state of dark energy over cosmic history.  If dark energy is a cosmological constant, then
$(w_0,w_a) = (-1,0)$.

The most commonly-used figure of merit $\mathcal{F}_{\rm DE}$
at present to characterise the
performance of a measurement of the properties of dark energy is the inverse area of the uncertainty
ellipse in the $(w_0,w_a)$ plane:
\begin{equation}
\qquad\qquad\qquad\qquad \mathcal{F}_{\rm DE} \; \equiv \; [{\rm det}\,\mathbf{C}(w_0,w_a)]^{-\frac{1}{2}},
\end{equation}
where $\mathbf{C}(w_0,w_a)$ is the covariance matrix of the two parameters~\citep{Alb06}.  Larger values of $\mathcal{F}_{\rm DE}$ represent
improved expected measurements, since larger values of $\mathcal{F}_{\rm DE}$ correspond to smaller uncertainties on the two dark
energy parameters $(w_0,w_a)$.

Both to generate simulated dataset catalogs of type~Ia supernovae that correspond to expected LSST observations
at the Rubin Observatory, and then to determine the best-fit cosmological parameter values and
associated uncertainties that result from those catalogs, we use the {\tt CosmoSIS} cosmological parameter estimation code~\citep{CosmoSIS}.
Approximately $40\,000$ SNeIa are expected to be observed in at least 4 filters per year in LSST~\citep{LSSTOverview,LSSTScienceBook}, and
thus we generate catalogs of $120\,000$ SNeIa each, in order to simulate the output of 3 years of Rubin Observatory operation.  For the generation of the
simulated catalogs, we utilise the same SALT2 parametrization of SNeIa that is used in the joint light-curve analysis (JLA) of
740 observed SNeIa~\citep{Betoule14}.  In particular, the generated distributions and correlations of the peak apparent magnitudes in the
rest-frame $B$ band $m_B^\star$, the generated observational redshifts $z$, the light-curve stretch factors $X_1$, and the colour
parameters $C$ of the simulated SNeIa, and the distributions of their respective uncertainties, within the catalogs are generated
according to the same distributions and correlations of those observational parameters that are found in the JLA.
Both the JLA, and our present analysis,
fit the data to a linear model whereby a standardised distance modulus $\mu \equiv 5 \log_{10}(d_{\rm L}/10\,{\rm pc})$ is
given by
\begin{equation}
\qquad\qquad\qquad\quad \mu \; = \; m_B^\star - (M_B - \alpha X_1 + \beta C),
\end{equation}
where $M_B$, $\alpha$, and $\beta$ are nuisance parameters which respectively correspond to the SNeIa absolute magnitude, and to light-curve
stretch and SNeIa colour correction factors to that absolute magnitude.  Also similarly to the JLA, and following the procedure previously
developed in~\citet{Con11}, we assume that the SNeIa absolute magnitude is related to the host galaxy stellar mass ($M_{\rm stellar}$) by a
step function:
\begin{equation}
\qquad\quad M_B \; = \; \left\{ \begin{array}{ll} M^1_B            & \quad {\rm if}\; M_{\rm stellar} < 10^{10} M_\odot ; \: {\rm and}\\
                                                  M^1_B + \Delta_M & \quad {\rm otherwise.} \end{array}
                         \right.
\end{equation}

However, the following modifications are made to the probability density distributions that are used for our catalog generation:
\begin{itemize}[label = \textbullet, leftmargin = 4mm, labelsep = *]
\item When generating the simulated SNeIa catalog that represents expected LSST observations \underline{\textbf{with}} LPRS-based
      photometric calibration, the generated systematic uncertainties on the SNeIa magnitudes are reduced by a factor of 2.18 from
      those in the JLA, corresponding to the expected improvement in SNeIa magnitude measurement from the LPRS-based photometric calibration;
\item Also when generating the simulated SNeIa catalog that represents expected LSST observations \underline{\textbf{with}} LPRS-based
      photometric calibration, the generated systematic covariances between the SNeIa magnitude and light-curve stretch values, as well
      as between the SNeIa magnitude and colour parameter values, are similarly reduced by a factor of 1.48 from those in the JLA,
      corresponding to the expected improvement in SNeIa magnitude measurement from LPRS-based photometric calibration.
\end{itemize}
No modifications to the distributions and correlations of the SALT2 parameters are made when generating the simulated SNeIa
catalog that represents expected LSST observations \underline{\textbf{without}} LPRS-based photometric calibration.
(I.e., the only statistical difference between the observed JLA SNeIa catalog and the simulated SNeIa catalog
representing expected LSST observations without LPRS-based photometric calibration is in the size of the catalog
itself: $120\,000$ SNeIa in the simulated catalog, vs.~740 SNeIa in the JLA observational
catalog.\footnote{This is not intended to be a perfect simulation of future Rubin Observatory SNeIa observations.  In particular, the redshift
distribution of the JLA SNeIa differs from the redshift distribution that is expected from LSST SNeIa.  Another important simplification is the
fact the we only generate and fit two simulated catalogs (instead of generating two large sets of simulated catalogs, with each member catalog
of each set containing $120\,000$ SNeIa, then fitting each member catalog of each set, and then considering the full distributions of the fitted
parameter results from each of the two sets of catalogs).  However, we note that the present paper is not intended to be centrally focused
on SNeIa catalog simulation and on fits thereof; but rather on the introduction of the concept of an LPRS, and on a simplified estimation of
the resulting impact on SNeIa dark energy measurements.})  The best-fit central value
of $w_0$ from a combined fit to the JLA data plus complementary probes is $-1.018$~\citep{Betoule14}, and thus we set $(w_0,w_a) = (-1.018, 0)$
for the generation of both of our two simulated SNeIa catalogs.

We fit the data within each of the two SNeIa catalogs for $\Omega_m$ (the matter density fraction of the critical density) ---
as well as for the four nuisance parameters $\alpha$, $\beta$, $M^1_B$, and $\Delta_M$ --- as free parameters in the fit,
which are the same 5 free parameters as in the nominal JLA data fit.  (And as in the nominal JLA data fit, we assume
a flat universe: the curvature parameter $\Omega_k = 0$, i.e.~the dark energy fraction of the critical
density $\Omega_{\rm DE} = 1 - \Omega_m$ in these fits.)  However,
within these fits to our two simulated catalogs, we also add the $w_0$ and $w_a$ variables mentioned above as additional free parameters
(whereas, in the nominal JLA fit, $w_0$ is fixed to $-1$, and $w_a$ is fixed to 0).

\begin{figure}
\begin{center}
\includegraphics[scale=0.52]{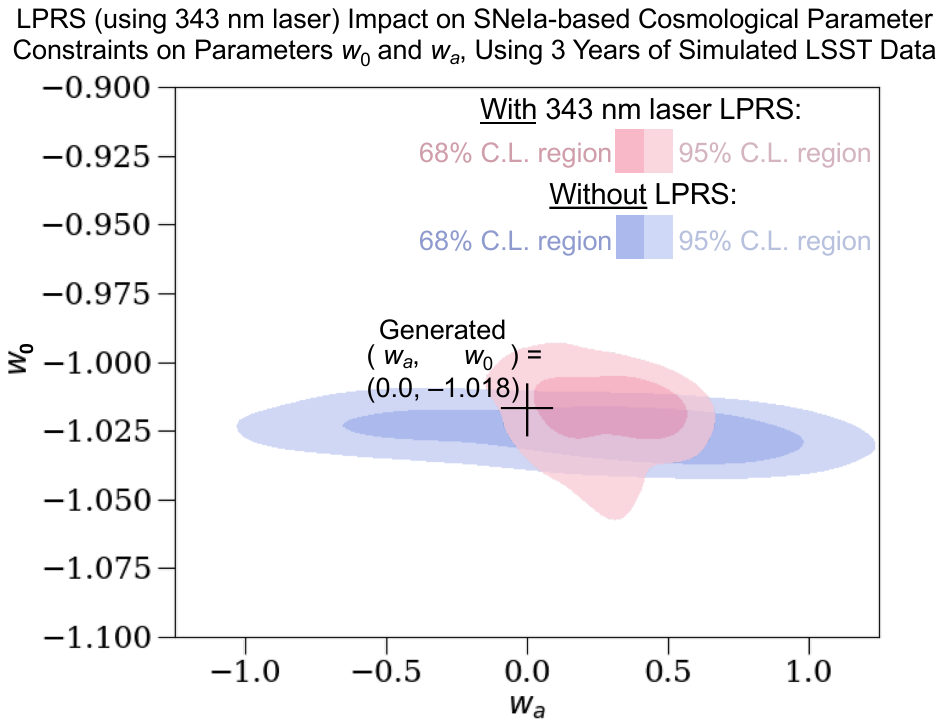}
\end{center}
\caption{Constraints on the dark energy equation of state parameters $w_0$ and $w_a$, obtained using
simulated catalogues of type~Ia supernovae.  Each one of the two SNeIa catalogues that are fitted to obtain
these constraints contains $120\,000$ simulated SNeIa (corresponding to approximately 3 years of observation at the Rubin Observatory).
The generation of the two simulated SNeIa catalogs, as well as the fits that are performed to each catalog, are
implemented using the {\tt CosmoSIS} cosmological parameter estimation code~\citep{CosmoSIS}, and are explained
in the text.
The resulting values of the figure of merit
parameter $\mathcal{F}_{\rm DE}$ for the fits are 313 for the fit to the SNeIa catalog representing expected LSST
observations without LPRS-based photometric calibration, and 607 for the fit to the SNeIa catalog representing
expected LSST observations with LPRS-based photometric calibration.
(Also, as would be expected on average, the central values of each fit are approximately one standard
deviation away from the generated values of $w_0$ and $w_a$.)
The fits represent a $\frac{607}{313} = 1.94$-fold
expected improvement in the dark energy figure of merit parameter $\mathcal{F}_{\rm DE}$ from LPRS-based photometric
calibration (with even greater expected
resulting $\mathcal{F}_{\rm DE}$ increases for SNeIa datasets that correspond to greater than 3 years of Rubin Observatory
observation).
}
\label{fig:LPRS1PhotImpactOnDECosmologyPlot}
\end{figure}

The results of the fits to the two simulated catalogs, when projected onto the $(w_0,w_a)$ plane (and, thus, when marginalised over
all of the other fitted parameters listed above), are shown in Figure~\ref{fig:LPRS1PhotImpactOnDECosmologyPlot}.
The resulting values of the figure of merit
parameter $\mathcal{F}_{\rm DE}$ for the fits are 313 for the fit to the SNeIa catalog representing expected Rubin Observatory
observations without LPRS-based photometric calibration, and 607 for the fit to the SNeIa catalog representing
expected Rubin Observatory observations with LPRS-based photometric calibration, representing a $\frac{607}{313} = 1.94$-fold
expected improvement in the dark energy figure of merit parameter $\mathcal{F}_{\rm DE}$ from LPRS-based photometric
calibration over 3 years of Rubin Observatory observation.  (Even larger resulting $\mathcal{F}_{\rm DE}$ increases due to this
LPRS-based photometric calibration would be expected for SNeIa datasets that correspond to greater than 3 years of  
Rubin Observatory observation.)

\section{Conclusions}

In summary, we present a method for establishing a precision reference for relative photometry between the visible and NIR
(and specifically between photometry at 589~nm and 819~nm wavelengths) using a powerful, mountaintop-located laser source tuned to the
342.78~nm vacuum excitation wavelength of neutral atoms of sodium.  As we have shown, if implemented this method would improve measurements   
of dark energy from type~Ia supernovae, using upcoming surveys such as the first 3 years of observations at the Vera C.~Rubin Observatory, by
approximately a factor of $1.94$ for the standard dark energy ``figure of merit'' $\mathcal{F}_{\rm DE}$ (which is based on the expected
uncertainties on measurements of the dark energy equation of state parameters $w_0$ and $w_a$).

This method could independently complement and cross-check other techniques under development for photometric calibration,
also of unprecedented
precision, that utilise laser diode and light-emitting diode light sources
carried on a small satellite~\citep{Albert12} and/or on small high-altitude
balloon payloads~\citep{Alb21b}, together with onboard precision-calibrated photodiodes
for real-time \textit{in situ} light source output measurement.

The utilization of order-of-magnitude improvement in the precision of
relative photometry between the visible and NIR will certainly not be limited only to SNeIa measurements of   
dark energy.  Within other areas of astronomy, precise measurements of stellar populations, photometric redshift surveys, and multiple other
astronomical measurements can benefit~\citep{Connor17,Kirk15}.  Even outside of astronomy, the use of
high-precision relative photometry with this technique could potentially help to pinpoint the quantities, types, and movement of aerosols
within the Earth's atmosphere above telescopes at night.

This technique could additionally be generalised to other atomic excitations, besides the 342.78~nm excitation of neutral sodium, that also
result in fully-mandated cascades consisting of specific ratios of de-excitation photons of different wavelengths.\footnote{In analogy with
LIDAR, the generalised technique could perhaps be termed LIDASP (LIght Detection And Spectro-Photometry).}  For example (although not so applicable within
the atmosphere of Earth), neutral hydrogen atoms have an analogous excitation wavelength of 102.57~nm in vacuum, which would result in a fully-mandated
cascade of 656.3~nm and 121.6~nm photons, which could possibly be used to explore, and to calibrate the exploration of, more
\ion{H}{i}-rich atmospheres of other planets or moons
when using a vacuum-UV laser source tuned to 102.57~nm.  Additionally, other neutral alkali metal atoms besides sodium have analogous
mandated-cascade-producing excitation wavelengths.

Excitations that result in fully-mandated cascades in the radio and microwave spectra, rather than in the UV, visible, or NIR spectra as in this paper, also
almost certainly exist, and thus could most likely be used for high-precision relative radiometry between wavelengths in those spectra, in an analogous manner
as with the optical photon technique that we describe.  Beyond even photometry and radiometry, the relative polarizations of the emitted
de-excitation photons in this technique will be anti-correlated with one another, and thus the precise calibration of relative polarimetry between
photons of wavelengths that equal those of the de-excitation photons from a cascade could be performed, especially if the photons emitted from the
source laser (or from the source maser) are of a definite and precisely-known polarization.

Thus, the technique described in this paper, as well as the related possible techniques mentioned in the above paragraphs in these Conclusions, have
prospects not only for SNeIa measurements of dark energy of unprecedented precision; but potentially in addition, more broadly,
for other novel measurements that utilise high-precision relative calibration of photometry, radiometry, or polarimetry,
both in astronomy and in atmospheric science, across the electromagnetic spectrum.

\section*{Acknowledgements}

The authors would like to thank Prof.~Gabriele Ferrari of Universit{\`a} di Trento and of LEOSolutions (Rovereto, Italy)
for critical and useful discussions regarding parametric crystal options for laser wavelength tunability; and Dr.~Torsten Mans
of Amphos GmbH (Aachen, Germany), Dr.~Knut Michel of TRUMPF GmbH (Ditzingen, Germany), and Dr.~Jochen Speiser of
the German Aerospace Center Institute of Technical Physics (Stuttgart, Germany) for further critical and useful
discussions regarding single-mode Yb:YAG disk pump lasers.  We would also like to thank Prof.~Chris Pritchet of the
University of Victoria for reading over the manuscript and providing extremely helpful comments and suggestions,
and Prof.~Christopher Stubbs of Harvard University for his helpful encouragement at an early stage of these papers.
JEA gratefully acknowledges support from Canadian Space Agency grants 19FAVICA28 and 17CCPVIC19.

\section*{Data Availability}

All code and data generated and used for the results of this paper is available from the authors upon request.
 



\bibliographystyle{mnras}
\bibliography{lidasp1} 





\bsp	
\label{lastpage}
\end{document}